\documentclass[12pt]{article}
\usepackage{cite}
\usepackage{amsmath}
\usepackage{amsfonts} \parindent 0pt \parskip.2cm \topmargin -1.0cm
\oddsidemargin=0.25cm\evensidemargin=0.25cm \newfont{\bbbold}{msbm10
  scaled \magstep1}

 \newcommand{\m}{{\mu}}

\def\pslash#1{{\setbox0=\hbox{$#1$}
    \rlap{\ifdim\wd0>.7em\kern.22\wd0\else\kern.1\wd0\fi /}#1}}

\def\be{\begin{equation}} \def\ee{\end{equation}}
\def\ba{\begin{array}} \def\ea{\end{array}}

\begin{document}

\thispagestyle{empty}
\begin{center}
{\Large\sf Conjugate variables in quantum field theory and a refinement of Paulis theorem }\\[10pt]
\vspace{.5 cm}

\vspace{1.0ex}
Steffen Pottel$^{\natural\dagger}$\footnote{e-mail: pottel`at'mis.mpg.de},
Klaus Sibold$^{\dagger}$\footnote{e-mail: sibold`at'physik.uni-leipzig.de}\\

$^{\natural}$Max-Planck Institute for Mathematics in the Sciences\\
Inselstr.\ 22\\
D-04103 Leipzig\\
Germany\\

$^{\dagger}$Institut f\"ur Theoretische Physik\\
Universi\"at Leipzig\\
Postfach 100920\\
D-04009 Leipzig, Germany

\end{center}

\vspace{2.0ex}
\begin{center}
{\bf Abstract}
\end{center}

\noindent
For the case of spin zero we construct conjugate pairs of operators
on Fock space. On states multiplied by polarization vectors 
coordinate operators $Q$ conjugate to the momentum operators $P$
exist. The massive case is derived from a geometrical quantity,
the massless case is realized by taking the limit $m^2\rightarrow 0$
on the one hand, on the other from conformal transformations.
Crucial is the norm problem of the states on which the $Q$'s 
act: they determine eventually how many independent conjugate
pairs exist. It is intriguing that (light-) wedge variables and hence
the wedge-local case seems to be preferred.\\

\newpage
\section{Introduction and embedding}
\subsection{Preliminaries}
\noindent
A student of physics meets conjugate pairs usually first in the context of
classical mechanics. Generalized coordinates $\{q_k\}_{k=1}^{k=n}$
serve together with generalized momenta $\{p_j\}_{j=1}^{j=n}$ as the
constitutive elements of Poisson brackets

\be\label{Poisson}
\{F,G\}=
\sum_{j,k}\left(\frac{\partial F}{\partial p^j}\frac {\partial G}{\partial q^k}
-\frac{\partial G}{\partial p^j}\frac{\partial F}{\partial q^k}\right)
\ee
for $F,G$ being functions of $p,q$ -- called observables. The $p$'s and
$q$'s span the phase space and the Poisson brackets define a symplectic
structure. Inserting for $F,G$ the momenta and coordinates themselves
one obtains

\be\label{classcp}
\{p_j,q_k\}=\delta_{jk}
\ee
with $\delta$ being Kroneckers $\delta$.
The (Hamiltonian) equations of motion read

\be\label{eqmot}
\frac{\partial H(p,q)}{\partial p^k}=\dot{q_k}\qquad
\frac{\partial H(p,q)}{\partial q^k}=-\dot{p_k}
\ee
with $H$ being the Hamiltonian of the system. The equation of motion for a
general
observable $O=O(p,q,t)$ which may explicitly depend on time is given by
the Poisson bracket

\be\label{eomPo}
\frac{dO}{dt}= \frac{\partial O}{\partial t} + \{H,O\}
\ee
The equations (\ref{eqmot}) become a case of (\ref{eomPo}) for $O=p_k$ and
$O=q_k$.
They are also known as {\sl canonical} equations of motion and transformations
$P=P(p,q), Q=Q(p,q)$ which leave them forminvariant are called canonical.
It is one of the beautiful results of classical mechanics that the actual
motion of a system in time i.e.\ the solutions of (\ref{eqmot})
$p_k(t),q_j(t)$, can be understood as a canonical transformation
which transports
initial data $p_k(t_0), q_j(t_0)$ to the actual one's at time $t$.\\
It is to be noted that time appears rather as a kind of ``external''
label than as coordinate. One may however incorporate it as $n+1$-th
coordinate and define -$H$ as its conjugate momentum \cite{Goldstein,FickI}.\\

\noindent
In relativistic point particle mechanics time becomes part of the coordinates
$x_\mu^{(j)}$ and may be re-introduced as {\sl eigentime} $\tau^{(j)}$
serving then as an {\sl invariant} for the labelling purpose along world lines
for the $j$-th particle.\\

\noindent
In quantum mechanics coordinates $q$ and momenta $p$ become Hermitian
operators $Q$ and $P$ acting on the state space of the system which is
a Hilbert space.
The Poisson brackets go (at least for Cartesian coordinates) over into
the commutator, the equations of motion (in the Heisenberg picture)
change accordingly

\be\label{ccrH}
[P_j,Q_k]=-i\delta_{jk}\quad\qquad i\frac{dO}{dt}=
              i\frac{\partial O}{\partial t}+[O,H]
\ee              
It is interesting to observe that the transformation $P\rightarrow -Q,
Q\rightarrow P$ is (like in classical mechanics) a canonical transformation,
which implies that if we choose as ``$q$-representation''
square integrable functions $f$ from, say, ${\mathbb{R}}^{3n} 
\rightarrow {\mathbb{C}}$ and consider their Fourier transforms (FT)
their role will be interchanged by the mentioned canonical transformation.
Realizing 
operators $P_j, X_k$ by the prescription

\begin{eqnarray}\label{Pj}
	P_j f(x)&=& -i\frac{\partial}{\partial x^j}f(x)
	    \qquad\mbox{FT}\qquad P_j \tilde{f}(p) = p_j\tilde{f}(p)\\
X_k f(x)&=& x_k f(x) \qquad\mbox{FT}\qquad X_k \tilde{f}(p)
	                  = i\frac{\partial}{\partial p^k}\tilde{f}(p)\\
\end{eqnarray}
(with j,k running from $1$ to $3n$), the roles of the operators will 
change accordingly. Invariant stay the relations

\begin{equation}\label{qm-com}
	 [P_j,X_k] = -i\delta_{jk},  \qquad [P_j,P_{j'}] = 0, \qquad [X_k,X_{k'}] = 0.\\
\end{equation}
The operators $P,Q$ are conjugate to each other and the FT indeed realizes
the conjugation.\\
The most intriguing aspect of the operator nature of observables is
certainly the discovery by Heisenberg that uncertainty relations hold
for observables which do not commute. Most notably in this context are
conjugate pairs.\\
Here $P_j$ generates translations in ${\mathbb{R}}^{3n}(x)$, whereas $X_k$
generates translations in ${\mathbb{R}}^{3n}(p)$. In quantum mechanics
the identification of $P_j$ with $3n$ momentum
operators and of $X_k$ with $3n$ position operators is automatic and the
unbounded operators $P_j$ and $X_k$ are essentially self-adjoint.
The role of Hamiltonian and an associated time operator is however 
special: the Hamiltonian is bounded from below, whereas a time operator
has to extend over the whole real line, hence
a tentative time operator cannot be self-adjoint. This is known as Pauli's
theorem \cite{Pauli} and precludes any naive extension to the relativistic 
situation.\\

\subsection{Embedding our approach}
The literature on position and time operators in quantum mechanics,
relativistic quantum mechanics and quantum field theory (QFT) is
overwhelmingly rich -- for a very good reason: the respective notions
are fundamental. We will not attempt to review it. Instead we
quote only a few papers to which our results may have a closer
relation. We regret all omissions.\\

\noindent
The impact of Poincar{\'e} invariance to the notion of localizability
in quantum theory has been analyzed in \cite{NewtonWigner}.
Under plausible assumptions on the set of states associated with
localization at a {\sl point} in three-dimensional space the authors arrive 
at the definition of a position
operator $x^{op}=i\nabla_{\bf p}-i{\bf p}/(2({\bf p}^2+m^2))$
acting on one-particle solutions of the Klein-Gordon equ.\ to mass $m$.
Thus, spatial localization at a point is not a Lorentz covariant concept.\\
In \cite{Wightman} the reference to a point in space has been weakened
to a {\sl finite region} in space, again quite plausible from a conceptional
point of view. The group theoretic analysis leads to the theorem that
all Lorentz invariant systems of $m^2>0$ are localizable and their position
variables are unique if the systems are elementary. For $m=0$, the only
localizable elementary system has spin zero.\\
\noindent
The next level of sophistication has been reached by {\sl local
quantum physics} in the spirit of \cite{Haag}. Over finite regions in
spacetime one defines nets of algebras of observables, studies their
representations and deduces their properties. A recent review of localization
based on these notions has been provided in \cite{Yngvason}. Quantum fields
may or may not be used in this context. It turns out that particle
states can never be created by operators strictly localized in bounded
regions of spacetime.\\
Our findings below better be in accordance with such general statements.\\
After this look into spatial localizability we should have a glance
at the construction of time operators.\\
Notable early papers are \cite{FickI,FickII,FickIII}. In analogy to classical
mechanics a time operator has been introduced and discussed within ordinary
quantum mechanics. It has been admitted as a Hermitian but not self-adjoint
operator. A wealth of further literature has been provided in \cite{Recami}.
On the more abstract level time operators are understood as
positive-operator-valued measures
\cite{BuschLahtiMittelstaedt,BuschGrabowskiLahti,BrunettiFredenhagenclick,BrunettiFredenhagentu}, 
or affiliated to $C^*$-algebras \cite{DoplicherFredenhagenRoberts}.
\noindent
A very recent review within the general context of quantum spacetime,
general relativity and even cosmology has been given in
\cite{BahnsDoplicherMorsellaPiacitelli}.\\

\noindent
For our own considerations reference to the role of the conformal
group is quite important. In \cite{Kastrup,KastrupMayer} the charges of
the special conformal transformations have become candidates for a
relativistic four position operator. From a different point of view
this has also been studied in \cite{HabaI,HabaII,HabaNowicki}. In detail
we will discuss below \cite{LaguLaue}. Eventually one would have to consider
the covering group $SU(2,2)$ which however is out of reach for the time
being. In \cite{PinamontiToller} the simpler case of $SU(1,1)$ has
been successfully treated and provides time observables with projective
covariance. Presumably, it this direction of research where to one
should find the connection with our treatment of the problem.\\
Our intention is to understand relativistic position operators as 
part of a theory which otherwise has been already constructed. 
Since models and their dynamics which are apt to experimental tests
rely even today mainly on perturbation theory the most
important Hilbert space for particle physics is Fock space and its imbedding
into systems of Green functions as off-shell continuation.
Available are conserved currents, their associated charges and composite
operators formed as functions of the basic quantum fields. Hence the most
useful tools are invariance groups and to some extent geometrical
quantities. Since in flat
spacetime Poincar\'e invariance is relevant the energy-momentum operator
$P$ participates in the game and a conjugate partner $Q$ is a natural
candidate for a position operator.\\
If one can dispose over conjugate pairs
one may define $Q'_\mu=Q_\mu+\Theta_{\mu\lambda}P^\lambda$ and obtains

\be\label{nccoord}
[Q'_\mu,Q'_\nu]=2i\Theta_{\mu\nu}
\ee
(for commuting $Q$'s). This relation is at the basis of some model classes
realizing non-commutative coordinates. This may provide additional motivation
for studying conjugate pairs in QFT. One may benefit in this context from
reading \cite{FredenhagenBulgJP} \footnote{We are grateful to Jochen Zahn
for pointing out this reference to us.}.\\
In \cite{Sibold,SiboldSolard,EdenSibold}
we have seen that it is non-trivial to realize the commutator

\be\label{bascom}
[P_\mu,Q_\nu]=i\eta_{\m\nu}
\ee
on Fock space immediately, that it is easier to study first {\sl prconjugate}
pairs $P,X$ which satisfy

\be\label{preconj}
[P_\mu,Q_\nu]=iN_{\m\nu},
\ee
where $N_{\mu\nu}$ is an operator which can (at least on states) be
inverted. In fact, previously we relaxed the diagonality condition expressed 
in the r.h.s.\ of (\ref{bascom}) which still yields interesting results
\cite{MuchPottelSibold}, but in the present paper we will study the
full strength of (\ref{bascom}) on states in Fock space and its surrounding
system of Green functions.\\
From all preconjugate operators introduced in \cite{MuchPottelSibold} we
will consider here in detail: $X(\nabla)$,\\ $X(<)$ and  $X(K)$ being based
on the mass shell belonging to four-dimensional Minkowski space and
$X(<_0),X(K)$ being based on $(1,1)+(0,2)$-dimensional spacetime.
The preconjugate $X(\omega)$ does not lead to Lorentz covariant $Q$ on 
$(1,3)$-dimensional spacetime and is therefore discarded. $X({\rm p-conf})$
turned out to be essentially $P$, hence does not need to be discussed.\\
Group theoretic considerations in section 3 serve to recapitulate earlier
work \cite{LaguLaue}, then to find a place for non-commutative coordinates, but
in particular -- via some new interpretation on
Fock space -- to control our derivations there. The distinguished role
played by
the special conformal generators as the only preconjugate $X$'s which are
local in position space  and permit a smooth transition between off-shell
and on-shell had been pointed out already in \cite{MuchPottelSibold}.
This explains why in the group theoretic context they have been singled
out.\\  
In section 4 we discuss our results, offer some conclusions and point out
open questions.\\

\newpage

\section{Conjugate operators in Fock space}
As mentioned already in the introduction we would like to construct operators $Q_\nu$ which act
in a sense to be specified as conjugate to the energy-momentum operator $P_\mu$ of the system:

\be\label{PQcom}
[P_\mu,Q_\nu]= i\eta_{\mu\nu}
\ee
On Fock space the right hand side of (\ref{PQcom}) cannot be a multiple of the
unit operator \footnote{KS is indebted to Rainer Verch for having pointed out 
to him the relevance of this fact.},
in particular, if $Q$ is charge like i.e.\ annihilates the vacuum, since $P$
does so by general assumptions
of QFT. Since we wish to obtain the $Q$'s also from charge like $X$'s we have
to understand
the commutator in a weak sense, namely applied to states -- here to states of
Fock space.
The definition of an appropriate $Q$ satisfying\\

\be\label{PQfock}
\lbrack P_\mu,Q_\nu\rbrack |{\bf p}_1,...{\bf p}_n> =
                            in\eta_{\mu\nu}|{\bf p}_1,...,{\bf p}_n>
\ee
 thus has to be found case by case.\\

\subsection{From $X(\nabla)$ to $Q(\nabla)$}
In \cite{MuchPottelSibold} we derived the operator

\be\label{Xnabla}
X^{(\nabla)}_\nu(a,a^\dagger)=\frac{i}{2}\int\!\frac{d^3p}{2\omega_p}\, 
        (a^\dagger({\bf p})\nabla_\nu a({\bf p})
       -\nabla_\nu a^\dagger({\bf p})a({\bf p})).
\ee                                
Here

\be\label{nabla}
\nabla_\nu\equiv\frac{\partial}{\partial p^\nu}
           -\frac{p_\nu p^\lambda}{m^2}
      \frac{\partial}{\partial p^\lambda}\quad \hbox{\rm with}\,
    \, p_0=\omega_p, \frac{\partial}{\partial p^0}=0\,  \hbox{\rm on-shell}.
\ee     
The operator $X^{(\nabla)}$ is charge like and (formally) Hermitian.\\
It satisfies the algebraic relation  

 \begin{align}\label{PXnablacom}
 [P_\mu, X^{(\nabla)}_\nu]&= i\int\!\frac{d^3p}{2\omega_p}(\eta_{\mu\nu}
 	            -\frac{p_\mu p_\nu}{m^2})a^\dagger({\bf p})a({\bf p})\\
		          &= i\eta_{\mu\nu}N-i\int\!\frac{d^3p}{2\omega_p}
		      \frac{p_\mu p_\nu}{m^2}a^\dagger({\bf p})a({\bf p}),
\end{align}
where $P_\mu$, $N$ denote the energy-momentum, resp.\ the number operator

\be\label{PNdef}
P_\mu = \int\!\frac{d^3p}{2\omega_p}\,p_\mu a^\dagger({\bf p})a({\bf p}),
     \qquad
         N = \int\!\frac{d^3p}{2\omega_p}\,a^\dagger({\bf p})a({\bf p}).
           \ee
We therefore qualified it as an operator {\sl preconjugate}
	   to $P$ on Fock space.\\
$X_\nu^{(\nabla)}$ transforms as a vector under Lorentz

\be\label{LonXnabla}
[M_{\mu\nu},X_\rho^{(\nabla)}] = i(X_\mu^{(\nabla)}\eta_{\nu\rho}
	                          -X_\nu^{(\nabla)}\eta_{\mu\rho})
\ee		
and for the commutator of $X$'s we found

\be\label{XXcom}
[X^{(\nabla)}_\mu,X^{(\nabla)}_\nu] =- \frac{i}{m^2}M_{\mu\nu}(a^\dagger,a)
\ee
On $n$-particle states $X^{(\nabla)}$ generates

\be\label{Xnablan}
iX_\nu^{(\nabla)}|{\bf p}_1,...,{\bf p}_n> =
                  \sum_{k=1}^n\left(\nabla_\nu^{(k)}-\frac{3}{2}p_\nu^{(k)}
                        \right)|{\bf p}_1,...,{\bf p}_n>
\ee
The aim is now to construct an operator $Q^{(\nabla)}_\nu$ such that it satisfies 

 \be\label{PQnablacom}
 [P_\mu, Q^{(\nabla)}_\nu] = i\eta_{\mu\nu}N
 \ee
on Fock space. Then we shall call this $Q$ conjugate to $P$.\\ 
In order to proceed we first apply (\ref{PXnablacom})
to the vacuum: the result is zero.\\
This originates from the fact that the operators involved are charge-like and
implements the aforementioned projector property of
the conjugation equation (\ref{PQcom}).\\
Applying (\ref{PXnablacom}) to an $n$-particle state yields

\be\label{nPXnablacom}
[P_\mu,X_\nu^{(\nabla)}]|{\bf p}_1,...,{\bf p}_n> = 
i\left(n\eta_{\mu\nu}-\sum_{k=1}^n\frac{p_\mu^{(k)}p_\nu^{(k)}}{m^2}\right)
                              |{\bf p}_1,...,{\bf p}_n>
\ee                              
This relation implies further projection content of (\ref{PQnablacom}):
for $n=1$ we have

\be\label{1PXnablacom}
[P_\mu,X_\nu^{(\nabla)}]|{\bf p}> = 
i\left(\eta_{\mu\nu}-\frac{p_\mu p_\nu}{m^2}\right)
                              |{\bf p}>
\ee                              
and obtain zero when contracting with $P^\mu$ from the left.
On states with $n>1$ the corresponding result is non-vanishing.
Furthermore, applying the commutator from the
l.h.s. of (\ref{1PXnablacom}) to a state $X^\nu(\nabla)|{\bf p}>$ and summing over $\nu$ we find

\be\label{projPPX}
[P_\mu,X_\nu^{(\nabla)}]X^\nu(\nabla)|{\bf p}>=i\nabla_\mu|{\bf p}>.
\ee
This relation can be read as $[P_\mu,X_\nu]$ being proportional to $i\eta_{\mu\nu}$ on
a ``non-trivial'' state -- a state $|{\bf p}>$ being multiplied by a non-trivial function of $p$.
This analysis, thus, suggests either to admit only states containing more than one particle
or to consider states
which are multiplied with non-trivial functions of the momenta. Let us study this latter case
 first.\\

 \subsubsection{Inversion on ``spin'' states}
 Since for $n=1$ the r.h.s.\ of \ref{1PXnablacom} is precisely the spin sum
 of a massive vector particle 

\be\label{spinsum}
\sum_{l=1}^3 \epsilon_\mu^{(l)}({\bf p)}\epsilon_\nu^{(l)}({\bf p})=
       -\left(\eta_{\mu\nu}-\frac{p_\mu p_\nu}{m^2}\right)
\ee
we are lead to introduce one-particle states

\be\label{spinstate}
|{\bf p},l,\mu>=\epsilon_\mu^{(l)}({\bf p})|{\bf p}>,
\ee
where 

\be\label{polariz}
\epsilon^{(l)}_\rho({\bf p})=
                           \left(\begin{array}{c}
		                  \frac{p^l}{m}\\
	                  -\delta^l_\rho+\frac{p^lp_\rho}{m(m+\omega_p)} \end{array}\right)
\ee
represents three $(l=1,2,3)$ polarization four-vectors, the first line giving
the $\rho=0$ component, whereas
the second line refers to their spatial components $\rho=1,2,3$.
\footnote{After having found (\ref{1PXnablacom}), recalled (\ref{spinsum})
	and then defined (\ref{spinstate}) the author KS understood a
	remark made to him earlier by Erhard Seiler, that the problem with
(\ref{PQcom}) is analogous to the state space problem in QED.} \\
They obey the orthogonality relations

\be\label{spinortho}
\epsilon^{(l')}_{\rho'}\eta^{\rho'\rho}\epsilon^{(l)}_\rho=\eta^{l'l}.
\ee

We first find

\be\label{spinnabla}
\sum_{l=1}^3\epsilon^{(l)}_\mu({\bf p)}\epsilon^{(l)}_\nu({\bf p)}\nabla^\nu =
                                                              - \nabla_\mu,
\ee
 and then 

\be\label{Xproj1}
i[X^\nu,[i[P_\mu,X_\nu],a^\dagger]]=\nabla_\mu a^\dagger,
\ee
(recall: $p_0=\omega_p, \partial/\partial p_0\equiv0 $); the contribution
$(3/2)p_\mu/m^2$ drops out. When applied to the vacuum state this means, that

\be\label{PXcomspin}
[P_\mu,X_\nu]\eta^{\nu\rho}\epsilon_\rho^{(l)}|{\bf p}>=i\epsilon_\mu^{(l)}|{\bf p}>.
\ee
I.e.\ the commutator operates on these states as $i\eta_{\mu\nu}$, which is
the desired conjugation relation on one-particle states. (A slightly different
way to derive (\ref{PXcomspin}) is to start from (\ref{nPXnablacom}) for $n=1$,
insert (\ref{spinsum}) in the r.h.s., to replace $|{\bf p}>$ by $|{\bf p},l,\mu>$ 
and then to use (\ref{spinortho})).\\
Due to the orthogonality relation (\ref{spinortho})
the vectors $|{\bf p},l,\mu>$ satisfy

\be\label{spinnorm}
<{\bf p}',l',\rho'|{\bf p},l,\rho> = 2\omega_p \delta({\bf p}'-{\bf p})
                 \epsilon^{(l')}_{\rho'}({\bf p})\epsilon^{(l)}_\rho({\bf p}),
\ee
hence have positive norm if we define their scalar product with the
metric $-\eta^{\rho'\rho}$.\\
An explicit form of operators $Q$ can be obtained as follows.
We consider 

\begin{align}\label{Xepsilona}
\eta^{\rho\sigma}[X_\rho,\epsilon^{(l)}_\sigma({\bf p})a^\dagger({\bf p})]
=& -i\eta^{\rho\sigma}\epsilon^{(l)}_\sigma\nabla_\rho a^\dagger({\bf p})\\
[X^\rho,\epsilon^{(l)}_\rho({\bf p})a^\dagger({\bf p})]
\doteq& -ie^{(l)}a^\dagger({\bf p})\\
e^{(l)}=& (-\delta^l_k+\frac{p^lp_k}{m(m+\omega_p)})
                               \frac{\partial}{\partial p_k}.
\end{align}            
These equations are all supposed to be applied to the vacuum, where
also for the commutator (\ref{PXcomspin}) the interchange
of the polarization vector with the operators $P,X$ is permitted.
Then the operators

\be\label{Qnabla}
Q^{(l)}_{\rm eff}|{\bf p}>= -ie^{(l)}|{\bf p}>
\ee
generate for ${\bf p}=0$, i.e.\ in the rest frame, precisely translations
in the momentum ${\bf p}$: they are indeed conjugate to $P$.
(We attached ``eff'' for ``effective'' because this equality only holds
when read in the context of (\ref{PXcomspin}).)\\
For finite, i.e.\ non-vanishing, ${\bf p}$ we use the fact that the
polarization vectors can be extended and then composed
to form a matrix $L$ with inverse $L^{-1}$

\begin{align}\label{boost}
	\left(L(p)\right)_\sigma^{\ \rho}= \left(\begin{array}{cc}
		\frac{\omega_p}{m}&-\frac{p^j}{m}\\
\frac{p_i}{m} & \delta_i^j-\frac{p_ip^j}{m(m+\omega_p)} \end{array}\right)
      \qquad    \hbox{\rm and} \qquad&
		\left(L^{-1}(p)\right)_\sigma^{\ \rho}= \left(\begin{array}{cc}
		\frac{\omega_p}{m}&\frac{p^j}{m}\\
\frac{-p_i}{m} & \delta_i^j-\frac{p_ip^j}{m(m+\omega_p)} \end{array}\right)
\end{align}
where $L$ is the boost, mapping the 4-vector $(m,0,0,0)^T$ into the
4-vector
$(\omega_p,p_1,p_2,p_3)^T$ and $L^{-1}$ transforms the derivatives

\be\label{boostinv}
(L^{-1}(p))_\sigma ^{\ \rho} \frac{\partial}{\partial p^\rho}=
      \left(\begin{array}{l}
		\frac{\omega_p}{m}\partial_0+\frac{p^j}{m}\partial_j\\
\frac{-p_i}{m}\partial_0+ \delta_i^j\partial_j-\frac{p_ip^j}{m(m+\omega_p)}\partial_j \end{array}\right).
\ee

\noindent
Since in the present context $\partial_0 \equiv 0$ we see first of all that
the contraction of $\epsilon$ with $\nabla$ results into the differential
operators $e$ in (\ref{Qnabla}). We may then go a step further and use the
fact that the first column of $L$ in (\ref{boost}) represents a fourth
timelike four vector $\epsilon^{(0)}_\rho$ which permits the definition

\begin{align}\label{Qeff}
Q^\lambda_{({\rm eff})}|{\bf p}>
	  =&X_\nu\eta^{\nu\rho}\epsilon^{(\lambda)}_\rho|{\bf p}>\\
          =&X_\nu\eta^{\nu\rho}(-L_\rho^{\,\lambda})|{\bf p}>\\
          =&-(L^{-1})^{\lambda\nu}X_\nu|{\bf p}>\\
Q^{({\rm eff})}_\lambda|{\bf p}>=&-i(L^{-1})_\lambda\,^\nu\nabla_\nu|{\bf p}>
\end{align}
(We have suppressed the contribution $\frac{3}{2}\frac{p_\nu}{m}$ within
	$X_\nu|{\bf p}>$ since it does not contribute eventually in the
	commutator $[P,X]$.)\\
Comparing with (\ref{boostinv}) we see that there we only have to replace the
ordinary by the tangential derivative to find the result

\begin{align}\label{Q0Qj}
	Q^{({\rm eff})}_0 |{\bf p}>=&0\\
	Q^{({\rm eff})}_j |{\bf p}>=&i\left(\frac{\partial}{\partial p^j} 
                       -\frac{p_jp^l\partial_l}{m(m+\omega_p)}\right)|{\bf p}>
\end{align}
For the commutators with $P_\mu$ this implies

\begin{align}\label{PmuQeff}
	[P_\mu,Q^{({\rm eff})}_0]|{\bf p}>=&0\\
	[P_\mu,Q^{({\rm eff})}_l]|{\bf p}>=&i\epsilon^{(l)}_\mu|{\bf p}>
                                    =-iL_\mu^{\, l}|{\bf p}>
\end{align} 
If we define

\be\label{Peff}
P^{({\rm eff})}_j=(L^{-1})_j^{\,\mu}P_\mu
\ee
we obtain finally

\begin{align}\label{PeffQeff}
	[P_\mu,Q^{({\rm eff})}_0]=[P^{({\rm eff})}_\mu,Q^{({\rm eff})}_0]=0\\
	[P^{({\rm eff})}_\mu,Q^{({\rm eff})}_l]|{\bf p}>=i\eta_{\mu l}|{\bf p}>
\end{align}

\noindent
As for the interpretation we may paraphrase the result as follows:
in the rest frame the polarization vectors are unit vectors and the
$X$'s coincide with the $Q$'s. As can be seen from (\ref{nabla}) at
${\bf p}=0\rightarrow\partial/\partial p^0=0$ in accordance with geometry: at
${\bf p}=0$ the tangential plane is orthogonal to the $p^0$-axis, hence
no tangential motion into that direction can be generated by an infinitesimal
change of ${\bf p}$. This implies $X_0=Q_0=0$.\\
At ${\bf p}=0$, the spatial $X$'s are conjugate
to the spatial $P$'s. For finite ${\bf p}$, we may with the help of 
polarization vectors define states with ``spin'' and introduce $Q$'s which
evolve with the inverse of these polarization vectors such that still
$Q_0=0$, the commutators with the $P$'s become polarization vectors,
which can then be absorbed into new $P$'s which are also just the evolved
one's for $P$. In this way the whole system remains Lorentz covariant.
The obvious analogue to this (from which the idea of introducing polarization
vectors has been suggested) is the quantization of a free, massive, abelian
vector field, s.\ \cite{Gasiorowicz}\cite{Sibold:1995sk}. There, like in the
present case, a
structure in three dimensional space is compatible with Lorentz covariance
in four dimensional spacetime by a correctly performed embedding: the
time component is a well determined function of  the space components.\\
Are the states $\epsilon|{\bf p}>$ asymptotic one's? Naively the answer
is ``yes'', since only on-shell momenta enter in their definition.
In $x$-space the polarization vectors represent non-local differential
operators, which can be seen e.g.\ when acting on a scalar field.
So, this may very well be an explicit realization of the general results
reported in \cite{Yngvason}.\\
Actually, already the operators $X^{(\nabla)}(a^\dagger,a)$ are
non-local when expressed in terms of the free scalar field,
in marked contrast to the conformal case, discussed below, since there
$X=K$ and the $K$'s are {\sl local} charges in $x$-space.
\footnote{$X(\nabla)$ represents the geometrical notion ``tangential derivative $\nabla$'' in Hilbert space, whereas $K$ represents the invariance of
$p^2=0$ there.}\\

\noindent
We still have to check how the commutator (\ref{XXcom}) translates itself
to the $Q$'s. It turns out that due to the presence of the polarization
vectors this commutator does not vanish. 
When searching for non-commutative coordinates one may thus rely on
preconjugate pairs \cite{MuchPottelSibold}, one may introduce
$\Theta$'s like in (\ref{nccoord}) or one can employ the operators
$Q^{({\rm eff})}$, (\ref{Q0Qj}). We hope to come back to this question
in the near future.\\

\noindent
For $n\ge 2$ one has to construct three four vectors which
are totally symmetric in the $n$ momenta, vanish when contracted with any one of them and are
reproduced by contraction with the transverse projector in the r.h.s. of (\ref{spinsum}). We do not
pursue this construction any further, since it is essentially provided by going over to the
helicity basis as used for scattering amplitudes.\\

\subsubsection{Inversion on standard states}   
We now wish to invert (\ref{PXnablacom}) on ordinary $n$-particle Fock states
in order to obtain effectively (\ref{PQcom}) on states. \\
By explicit calculation we find

\be\label{commnabla}
[P_\mu,X^{(\nabla)}_\nu]|{\bf p}_1,...,{\bf p}_n>=
i\sum_{k=1}^n (\eta_{\mu\nu}-\frac{p_\mu^{(k)} p_\nu^{(k)}}{m^2})|{\bf p}_1,...,{\bf p}_n>,
\ee
and the question is, whether the $4\times 4$-matrix (in the indices $\mu,\nu$) is invertible.
As noted above this is not the case for $n=1$, since $P^\mu$ projects to zero. For $n=2$
one checks in the center-of-mass system ${\bf p}\equiv{\bf p}_1=-{\bf p}_2$ that the determinant results
into

\be\label{detm}
det(\mbox{r.h.s})=-\frac{16}{m^4}{\bf p}^2\omega^2_p\not=0.
\ee
Hence this matrix can be inverted, the inverse applied from the right and attributed as factor to
$X$, which thereby becomes a $Q$. (The momentum ${\bf p}=0$ is an unphysical point.)\\
Since for $n$ larger than two the kinematical configuration can not become worse, we conclude
that the inversion is possible for all $n\ge 2$.\\ 
Let us now discuss the case $n=2$ in more detail. (\ref{commnabla}) reads 
\begin{align}\label{commnabla2}
[P_\mu,X^{(\nabla)}_\nu]|{\bf p}_1,{\bf p}_2>=& 2iN_{\mu\nu}|{\bf p}_1,{\bf p}_2>\\
                 \hbox{\rm with}\quad N_{\mu\nu}=&\left(\eta_{\mu\nu}-\frac{p_\mu^{(1)} p_\nu^{(1)}}{2m^2}
-\frac{p_\mu^{(2)}p_\nu^{(2)}}{2m^2}\right)
\end{align}
In the center-of-mass system and after rotating to zero the $y$- and $z$-components
of ${\bf p}$ the matrix $N_{\mu\nu}$ is diagonal

\be\label{diagN}
N_{\mu\nu}=-\left(\begin{array}{cccc}
		\frac{p_{x}^2}{m^2}&0&0&0\\
		0&\frac{2(m^2+p_{x}^2)}{m^2}&0&0\\
                       0&0&1&0\\
                       0&0&0&1
                             \end{array}\right),
\ee
Multiplying (\ref{commnabla2}) with the inverse of $N$ 

\be\label{2nablainv}
{(N^{-1})}^{\nu\rho}=\left(\begin{array}{cccc}
                    -\frac{m^2}{p_x^2}&0&0&0\\
		0&\frac{m^2}{2(m^2+p_x^2)}&0&0\\
                    0&0&1&0\\
                    0&0&0&1
\end{array}\right)^{\nu\rho},
\ee
We arrive at the conjugation equation in the form

\begin{align}\label{conjeqnabla2}
	[P_\mu,Q_\nu]|{\bf p},-{\bf p}>\, =& \, 2i\eta_{\mu\nu}|{\bf p},-{\bf p}>\\
	Q_0=-\frac{m^2}{p_x^2}X_0\quad & \quad Q_1=\frac{m^2}{2(m^2+p_x^2)}X_1\\
	Q_2=X_2\quad & \quad  Q_3=X_3
\end{align}	
Like in the preceding subsubsection we have to check now the norms of the
states created by the commutator $[P_\mu,Q_\nu]$. The one generated by
$[P_0,Q_0]$ is opposite to the one generated by the spatial components
$[P_j,Q_j]$ ($j=1,2,3$ no sum). Hence we face a problem which is just
the same one faces in gauge theories: the scalar component
$\partial_\mu A^\mu$ of the vector field creates states with negative
norm. Thus we try to remedy it by the same mean as there: we impose
a Gupta-Bleuler condition on the allowed {\sl states}, thereby characterizing
them as physical ones. Combining the contribution from the $(0,0)$-component
with that of the $(1,1)$ component and requiring that the sum vanishes
we find

\be\label{GuptaBleulerm}
\left(-\frac{2m^2}{p_x^2}\alpha_0+\frac{m^2}{m^2+p_x^2}\alpha_1\right)
|{\bf p},-{\bf p}>=0.
\ee
(Here the $\alpha$'s are real numbers.) This equation has no solution
identically in {\bf p}. However in the massless
limit such a solution exists with $\alpha_1=2\alpha_0$.\\
We conclude from this result that in the massive case such an inversion procedure
is not consistent. Only the construction of the preceding
subsubsection seems to be applicable.\\
Let us have a look at the massless limit. Obviously $Q_0=Q_1=0$. This
tells us that only the spatial components $Q_2$ and $Q_3$ exist and are
conjugate to $P_2, P_3$ respectively. Effectively, the measurable quantities
are these spatial ones. Hence this solution is not {\sl manifestly}
Lorentz covariant, but nevertheless covariant in the sense of the transition
from $\{P,Q\}$ to $\{P_{\rm eff},Q_{\rm eff}\}$ above and the case of
$Q(K)$ treated below in section 2.3.\\

\subsection{From $X(<_0)$ to $Q(<_0)$} 
In \cite{MuchPottelSibold} we introduced wedge variables (``$<$'' for 
``wedge'')      

\begin{eqnarray}\label{wedgevariables}
	\mbox{in $p$-space}\qquad	        p_u&=&\frac{1}{\sqrt{2}}(p_0-p_1) \qquad\qquad p_0=\frac{1}{\sqrt{2}}(p_v+p_u)\\
	        p_v&=&\frac{1}{\sqrt{2}}(p_0+p_1) \qquad\qquad p_1=\frac{1}{\sqrt{2}}(p_v-p_u)\\
	\mbox{in $x$-space}\qquad	        u&=&\frac{1}{\sqrt{2}}(x^0-x^1) \qquad\qquad x^0=\frac{1}{\sqrt{2}}(v+u)\\
	      v&=&\frac{1}{\sqrt{2}}(x^0+x^1) \qquad\qquad x^1=\frac{1}{\sqrt{2}}(v-u)
\end{eqnarray}
Note: $p^u=p_v, p^v=p_u$.
The mass shell condition is given by

\be\label{wedgemassshell}
2p_up_v-p_ap_a = m^2 \qquad a=2,3 \qquad\qquad{\mbox{summation over a}}
\ee
We then constructed differential operators
$\nabla(<)_\mu, \quad \mu=u,v,2,3$ acting on one-particle wave functions. Here
one can admit $p_0=\pm \omega_p, \quad \omega_p\equiv \sqrt{m^2+{\bf p}^2}$,
hence {\sl both} shells of the hyperboloid $p^2=m^2$ are covered. When aiming at
operators $X(a,a^\dagger)$
for realizing these differential operators on Fock states one has to introduce
new creation and annihilation operators, since the standard one's are based
on $p_0=+\omega_p$.\\
One can proceed as follows \cite{HeslopSibold}. A scalar field satisfying
the Klein-Gordon equation is being introduced as

\be\label{wedgefield}
\phi(x)=\frac{1}{(2\pi)^{3/2}}\int\frac{d^3p}{2p_v}e^{-i{\bar p}x}A({\bf p})
\ee
with $d^3p\equiv dp_vdp_2dp_3, {\bf p}=(p_v, p_a),
\bar{p}_u=(m^2+p_ap_a)/(2p_v), \bar{p}_v=p_v, \bar{p}_a=p_a\, a=2,3$.
Reality of $\phi$ implies

\be\label{real}
A^\dagger({\bf p})=-A({\bf -p)}
\ee
One can invert (\ref{wedgefield})

\be\label{Aofphi}
A({\bf p})=\frac{1}{(2\pi)^{2/3}}\int d^3x 2p_v e^{i\bar{p}x}\phi(x)
\ee
The field is quantized by imposing

\begin{align}\label{ccrwedge}
		[A({\bf p}),A({\bf p'})]=&\,2p_v\delta^3({\bf p}+{\bf p'})\\
	A({\bf p})|0>=0 \quad {\rm for}\, p_v<0 \qquad & \quad
	 <0|A({\bf p})=0 \quad {\rm for}\, p_v>0
\end{align}
Below we shall need this definition of Fock states because it will serve
to clarify the relations amongst the different $Q$'s which we study.
For the purposes of the present discussion we work however with the
differential operators for which the
respective modifications are essentially trivial.\\
We treat here $\nabla(<_0)$ and discuss in terms of it the
properties of $X(<_0)$ and $Q(<_0)$). We found in \cite{MuchPottelSibold}

\be\label{wedge0def}
\nabla^u=\frac{1}{2}(\frac{\partial}{\partial p_u}-\frac{1}{p_u}p_v\frac{\partial}{\partial p_v})\qquad
\nabla^v=\frac{1}{2}(\frac{\partial}{\partial p_v}-\frac{1}{p_v}p_u\frac{\partial}{\partial p_u})
\ee
\be
\nabla^2 = \frac{\partial}{\partial p_2} -\frac{p^2}{p_ap^a}p_b\frac{\partial}{\partial p_b} \qquad
\nabla^3 = \frac{\partial}{\partial p_3} -\frac{p^3}{p_ap^a}p_b\frac{\partial}{\partial p_b}.
\ee
These differential operators satisfy the algebra

\begin{align}\label{)uvalgeb}
[\nabla^u,\nabla^v]&=\frac{-1}{2p_up_v}(p_u\frac{\partial}{\partial p^v} 
                        -p_v\frac{\partial}{\partial p^u})
                    =\frac{1}{p_ap^a}(p_u\frac{\partial}{\partial p^v} 
                        -p_v\frac{\partial}{\partial p^u})\\
[\nabla^u,\nabla^2]&=[\nabla^u,\nabla^3]=[\nabla^v,\nabla^2]=[\nabla^v,\nabla^3]=0\\
[\nabla^2,\nabla^3]&=-\frac{1}{p_ap^a}(p^2\frac{\partial}{\partial p_3}
			-p^3\frac{\partial}{\partial p_2})
                    =\frac{1}{2p_up_v}(p^2\frac{\partial}{\partial p_3}
                        -p^3\frac{\partial}{\partial p_2}).          
\end{align}
They furthermore obey projection properties

\be\label{lwedgeproject}
p_u\nabla^u+p_v\nabla^v =0 \qquad\qquad p_2\nabla^2+p_3\nabla^3=0.
\ee
We defined accordingly operators $X(<_0)$ acting on functions
$\tilde{f}(p_u,p_v,p_2p_3)$ as differential operators by
\begin{eqnarray}
	                X^u(<_0)&=&i\nabla^u  \qquad X^2(<_0)=i\nabla^2\\
	        X^v(<_0)&=&i\nabla^v  \qquad X^3(<_0)=i\nabla^3.
\end{eqnarray}
Their algebra is given by
\begin{align}
[X^u(<_0),X^v(<_0)]&=i\frac{1}{P_aP^a}M^{uv}\\
[X^2(<_0),X^3(<_0)]&=i\frac{1}{2P_uP_v}M^{23}\\
[X^u(<_0),X^2(<_0)]&=[X^u(<_0),X^3(<_0)]=[X^v(<_0),X^2(<_0)]=[X^v(<_0),X^3(<_0)]=0.
\end{align}
Their commutation relations with the energy-momentum operator $P$ read

\begin{align}\label{lightwPXcom}
[P_\alpha,X_\beta(<_0)]&=\frac{i}{2}\left(\begin{array}{cc}
	                         -\frac{P_u}{P_v} &1\\
                        	                   1 &-\frac{P_v}{P_u}
		  \end{array}\right)_{\alpha\beta}    \quad \alpha,\beta=u,v \\
[P_a,X_b(<_0)]&=-i\left(\begin{array}{cc}
                        1+\frac{P_2P_2}{P_bp^b}&\frac{-P_2P_3}{P_bP^b}\\
                        \frac{-P_3P_2}{P_bP^b} &1+\frac{P_3P_3}{P_bP^b}
	  \end{array}\right)_{ab}  \quad a,b=2,3
\end{align}
The main implication of this structure is the loss of symmetry: from the
original $SO(1,3)$ invariance survived only $SO(1,1)\times SO(2)$. The
remaining generators do not exist in the limit of vanishing mass and have
thus to be excluded from participation. This is to be compared with the limit
$m^2=0$ taken at the end of the preceding subsection: there no boost
survived -- the limit was effectively non-relativistic, although Lorentz
covariance was not lost.\\
Some more information from this limit process will be useful later on.
Using the transformation equations (\ref{wedgevariables}) we find that

\begin{align}\label{limwedgediff}
\mbox{for}\qquad  p_0=+p_1>0&\qquad \nabla^u \quad\mbox{does not exist}\\
            &\qquad \nabla^v=\frac{1}{2\sqrt{2}}(\partial_0-\partial_1)\\
\mbox{for}\qquad  p_0=-p_1>0&\qquad \nabla^v \quad\mbox{does not exist}\\
		&\qquad \nabla^u=\frac{1}{2\sqrt{2}}(\partial_0+\partial_1)
\end{align}

\subsubsection{The $SO(2)$-sector}
In close analogy to the massive case we try to realize the conjugation
structure on states multiplied by polarization vectors. We choose

\be\label{pol0vec}
\epsilon^{(2)}_a=\frac{1}{|{\bf p}|}\left(\begin{array}{c}
		|{\bf p}|\cos\alpha\\
                |{\bf p}|\sin\alpha \end{array} \right) \qquad
\epsilon^{(3)}_a=\frac{1}{|{\bf p}|}\left(\begin{array}{c}
		-|{\bf p}|\sin\alpha\\
                |{\bf p}|\cos\alpha  \end{array}\right)
		\qquad a=2,3 \qquad |{\bf p}|\equiv \sqrt{p_2^2+p_3^2}
\ee	
An equivalent form is

\be\label{pol0vec'}
\epsilon^{(2)}_a=\frac{1}{|{\bf p}|}\left(\begin{array}{c}
		 p_2\\
                p_3 \end{array} \right) \qquad
\epsilon^{(3)}_a=\frac{1}{|{\bf p}|}\left(\begin{array}{c}
		-p_3\\
                p_2  \end{array}\right) \qquad a=2,3
\ee	
with the obvious identification $p_2=|{\bf p}|\cos\alpha, p_3=|{\bf p}|\sin\alpha$.
They are spacelike unit vectors

\be\label{pol0unit}
\epsilon^{(2)}_a\eta^{ab}\epsilon^{(2)}_b=
\epsilon^{(3)}_a\eta^{ab}\epsilon^{(3)}_b=-1
\ee
and satisfy the completeness relation

\be\label{p0compl}
\epsilon^{(2)}_a\epsilon^{(2)}_b+\epsilon^{(3)}_a\epsilon^{(3)}_b=
 \left(\begin{array}{c}
 		 1\quad 0\\
                 0\quad 1
       \end{array}\right) =-\eta_{ab}
\ee       
The right hand side of (\ref{lightwPXcom}) indeed is then equal to
$-i\sum_c^{2,3}\epsilon^{(c)}_a\epsilon^{(c)}_b$ and we may expect that

\be\label {Xw0p}
X_b\eta^{bc}\epsilon^{(r)}_c|{\bf p}>=\eta^{bc}\epsilon^{(r)}_ci\nabla_b|{\bf p}>
\ee
gives rise to an effective conjugate:

\be\label{Qwso2}
Q^{(r)}_{\rm eff}|{\bf p}>=i\eta^{bc}\epsilon ^{(r)}_c\nabla_b|{\bf p}>
\ee 
The explicit calculation leads to

\begin{align}
Q^{(2)}_{\rm eff}|{\bf p}>=&0\\
Q^{(3)}_{\rm eff}|{\bf p}>=&\frac{i}{|{\bf p}|}(-p_3\frac{\partial}{\partial p_2}
	+p_2\frac{\partial}{\partial p_3})|{\bf p}>.
\end{align}
For the effective commutator with $P$ this implies

\begin{align}\label{PQlwso2}
	[P_2,Q^{(3)}_{\rm eff}]|{\bf p}>=&-i\sin\alpha|{\bf p}>\\
	[P_3,Q^{(3)}_{\rm eff}]|{\bf p}>=&i\cos\alpha|{\bf p}>
\end{align}
Therefore the system has {\sl one} independent conjugate pair which corresponds to the
fact, that 
the commutator matrix (\ref{lightwPXcom}) has vanishing determinant which in turn
originates from the projector property (\ref{lwedgeproject}).\\

\noindent
The normalization properties (\ref{pol0unit}) tell us that the states
$\epsilon {(r)}|{\bf p}> \qquad r=2,3$ have positive norm, if we introduce
the metric $\eta_{rs}\quad r,s=2,3$ in this state space.\\

\subsubsection{The $SO(1,1)$-sector}
Similarly to the choice of polar variables in the previous subsubsection it turns out
that in the present sector hyperbolic variables are most suitable. We introduce

\begin{align}\label{hypvar}
	p_u=\frac{c}{\sqrt{2}}(\cosh\phi-\sinh\phi)=\frac{c}{\sqrt{2}}e^{-\phi}
	&\quad c=\sqrt{2p_up_v}=\sqrt{p_ap_a} \quad {\rm sum}\quad a=2,3\\
	p_v=\frac{c}{\sqrt{2}}(\cosh\phi+\sinh\phi)=\frac{c}{\sqrt{2}}e^{+\phi}
	&\quad \phi=-\frac{1}{2}\ln\frac{p_u}{p_v}=\frac{1}{2}\ln\frac{p_v}{p_u}
\end{align}	
(Note: since always $p_up_v>0$ the functions involved are well-defined.)\\
The commutator matrix (\ref{lightwPXcom}) assumes the form

\be\label{hypPXcom}
[P_\alpha,X_\beta(<_0)]=
\frac{i}{2}\left(\begin{array}{cc}
		-\frac{p_u}{p_v} & 1\\
1 &-\frac{p_v}{p_u}\end{array}\right)_{\alpha\beta}
= \frac{i}{2}\left(\begin{array}{cc}
		-e^{-2\phi} & 1\\
1 & -e^{2\phi}\end{array}\right)_{\alpha\beta}
\ee
The tangential derivatives $\nabla^u, \nabla^v$ applied to a one-particle state

\be\label{opstuv}
|{\bf p}>=|p_u,p_v;p_2,p_3>_{|p_u=\frac{p_ap_a}{2p_v}}\, \equiv |\cdots>
\ee
become

\be\label{hyptander}
\nabla^u|\cdots>=
-\frac{1}{\sqrt{2}}\frac{e^\phi}{c}\frac{\partial}{\partial \phi}|\cdots>
\qquad \nabla^v|\cdots>=
\frac{1}{\sqrt{2}}\frac{e^{-\phi}}{c}\frac{\partial}{\partial \phi}|\cdots>
\ee
Geometrically interpreted this means that they generate motions on the
hyperbolae $p_u=p_u(\phi), p_v=p_v(\phi)$ for fixed $c=\sqrt{p_ap_a}$.
Their projection properties (\ref{lwedgeproject}) are, of course, maintained.\\

\noindent
We now introduce polarization vectors

\be\label{poluvvect}
\epsilon^{(u)}_\alpha =
          \frac{N_u}{\sqrt{2}}\left(\begin{array}{cc}
			  -\frac{p_u}{p_v}\\     
                               1 \end{array}\right) \qquad
\epsilon^{(v)}_\alpha =
          \frac{N_v}{\sqrt{2}}\left(\begin{array}{cc}
			            1\\
			  -\frac{p_v}{p_u} \end{array}\right),
\ee
where $N_u,N_v$ are arbitrary normalization factors, calculate
their normalization

\be\label{normepsuv}
\epsilon^{(\sigma)}_\gamma\eta^{\gamma\beta}\epsilon^{(\tau)}_\beta
=\left\{\begin{array}{cc}
-\frac{p_u}{p_v}N^2_u&\quad {\rm for}\quad \sigma=u, \tau=u \\
	N_uN_v &\quad {\rm for}\quad \sigma=u, \tau=v\\
        N_uN_v &\quad {\rm for}\quad \sigma=v, \tau=u\\
-\frac{p_v}{p_u}N^2_v&\quad {\rm for}\quad \sigma=v, \tau=v
\end{array}\right.	
\ee
and their completeness relation

\be\label{poluvcompl}
\sum_{\tau}\epsilon^{(\tau)}_\alpha\epsilon^{(\tau)}_\beta=
               \frac{N_u^2p_u^2+N_v^2p_v^2}{2p_up_v}
                         \left(\begin{array}{cc}
                          \frac{p_u}{p_v}&-1\\
		          -1&\frac{p_v}{p_u}
			 \end{array}\right)_{\alpha\beta}
\ee
For the commutator (\ref{hypPXcom}) we can therefore write

\be\label{uvcomPX<}
[P_\alpha,X(<_0)_\beta]|\cdots>
        =i\sum_{\tau}\epsilon^{(\tau)}_\alpha\epsilon^{(\tau)}_\beta
           \frac{p_up_v}{N^2_up^2_u+N^2_vp^2_v}|\cdots>
\ee
Applying this commutator to the state
$\eta^{\beta\gamma}\epsilon^{(\sigma)}_\gamma|\cdots>$ we find by
explicit calculation the expected result, namely

\be\label{uvPQ}
[P_\alpha,X(<_0)_\beta]\eta^{\beta\gamma}\epsilon^{(\sigma)}_\gamma|\cdots>
=-i\delta_\alpha^\gamma\epsilon^{(\sigma)}_\gamma|\cdots>,
\ee
i.e.\ the l.h.s.\ acts as a $[P_\alpha,Q^\gamma_{\rm eff}]$ on these
states, with

\be\label{Qexpl}
Q^u_{\rm eff}=-i\frac{N_u}{\sqrt{2}}\nabla^v \qquad
Q^v_{\rm eff}=-i\frac{N_v}{\sqrt{2}}\nabla^u 
\ee
Hence we have two pairs of conjugate operators. This is due to the fact
that the singularity  for vanishing $p_u,p_v$ prohibits the transition
from the upper part of the hyperboloid to the lower one (and vice versa)
and that the respective reflection is not in $SO(1,1)$.\\
If we now choose $N_u=N_v\equiv N$ which is possible (e.g. with
$N=\sqrt{\frac{p_u^2+p_v^2}{2p_up_v}}$), then the $Q^\sigma_{\rm eff}$
operate just like a rescaled $X^\sigma$ (on different states however),
hence transform as a vector under $SO(1,1)$ and have a non-trivial
commutator

\be\label{Lorgen}
[Q^\sigma_{\rm eff},Q^\tau_{\rm eff}]|\cdots>=
        -\frac{1}{p_ap^a}(p^\sigma\frac{\partial}{\partial p_\tau}
            -p^\tau\frac{\partial}{\partial p_\sigma})  
                   \equiv\frac{i}{P_aP^a}M^{\sigma\tau}|\cdots>,
\ee
with $M$ being the generator of $SO(1,1)$. From $\nabla^u, \nabla^v$
they inherit on the states $|\cdots>$ the functional dependence

\be\label{Qdep}
(p_uQ^u_{\rm eff}+p_vQ^v_{\rm eff})|\cdots>=0
\ee
Reading the equations (\ref{Qexpl},\ref{Lorgen}) in terms of the hyperbolic
variables (\ref{hypvar},\ref{hyptander}) we have a perfect analogy to the
purely spatial sector  with its covariance under the compact group $SO(2)$.

\noindent
The norms of the states
$\epsilon^{(\sigma)}_\alpha|p_u,p_v;p_2,p_3>_{|p_u=\frac{p_ap_a}{2p_v}},
\sigma\in\{u,v\}$ can be read off from (\ref{normepsuv}) for $N_u=N_v\equiv N$
and are positive definite for $\sigma=u,v$ respectively:

\begin{align}\label{normuvsts}
<{\bf q}|\epsilon^{(\sigma)}_\alpha({\bf q})\eta^{\alpha\beta}
                \epsilon^{(\sigma)}_\beta({\bf p})|{\bf p}>
		=&N^2\delta^{(2)}(q-p)\delta(p_v+q_v)
		<q_v;q_2,q_3|p_v;p_2,p_3>\times\nonumber\\ 
			   &\times\left\{\begin{array}{c}
		   \frac{p_u}{p_v} \quad {\rm for}\, \sigma=u\\
		   \frac{p_v}{p_u} \quad {\rm for}\, \sigma=v
				\end{array}\right.
\end{align}                            
If we introduce the metric $\eta^{\alpha\beta}$ we have a positive definite
norm for these states, maintaining covariance.\\

\noindent
The extension from the one-particle situation to $n$ particles by tensoring
deserves further study: the introduction of relative momenta and separation
of the center-of-mass Hamiltonian as it has been studied in light-cone
quantization (s.\ \cite{Heinzl} for a comprehensive review)
should be complemented by the analogous treatment of the $q$-variables
and could yield quite interesting results.\\

\subsubsection{From $X(<)$ to $Q(<)$}
Having discussed the massless case $<_0$ and seeing no obvious reason
why the extension to the massive case should not work we establish
now the analogous structure there. For the case $<_0$ the relevant spacetime
was $(1,1)\times (0,2)$ having the symmetry $SO(1,1)\times SO(2)$. In the
massive case one can realize manifestly also first this symmetry, discuss the
construction of $Q$'s and state space associated with it and thereafter
implement the symmetry generators missing to the complete $SO(1,3)$.\\
This can be seen as follows.
Again we base our analysis on the differential operators and not on the
Fock space expressions, since the difference between the two versions
can safely be expected to be a contribution proportional to $P_\mu$, hence
not contributing to the commutator $[P,Q]$.\\
In \cite{MuchPottelSibold} we found differential operators $\nabla(<)$
tangential to the mass shell $2p_up_v=m^2+p_ap_a$

\begin{eqnarray}\label{tangderans}
	\nabla^u = \frac{\partial}{\partial p_u} - \frac{p_v}{m^2}p_\lambda\frac{\partial}{\partial p_\lambda}
	&\qquad&
	\nabla^2 = \frac{\partial}{\partial p_2} + \frac{p^2}{m^2}p_\lambda\frac{\partial}{\partial p_\lambda}\\        
	\nabla^v = \frac{\partial}{\partial p_v} - \frac{p_u}{m^2}p_\lambda\frac{\partial}{\partial p_\lambda}
	&\qquad&
	\nabla^3 = \frac{\partial}{\partial p_3} + \frac{p^3}{m^2}p_\lambda\frac{\partial}{\partial p_\lambda},
\end{eqnarray}
which gave rise to operators

\be\label{mXwedge}
X(<)=i\nabla(<),
\ee
with commutator

\be\label{mPXwedgecom}
[P_\mu,X_\nu^{(<)}]f(p) = 
i\left(\bar{\eta}_{\mu\nu}-\frac{p_\mu p_\nu}{m^2}\right)f(p).
\ee                              
Here the indices $\m,\nu$ run over the ranges $\{u,v,2,3\}$ and the metric
$\bar\eta$ reads

\be\label{wedgemetric}
\bar{\eta}_{\mu\nu}=\left(\begin{array}{cccc}
		                          0&1&0&0\\  
		                          1&0&0&0\\
		                          0&0&-1&0\\
		                          0&0&0&-1\\ \end{array}\right).
\ee
The functions $f$ stand for eigenfuctions of the energy-momentum operator
in terms of the wedge variables $p$, thus permitting the transition to the
mass-shell accordingly.\\
It is now crucial to observe that in the massive case a partial rest system
with $p_2=p_3=0$ exists in which the $\{2,3\}$-sector is diagonal, whereas
the $\{u,v\}$-sector assumes the form

\be\label{mrPXcom}
[P_\alpha,X^{(<)}_\beta]f(p)=-i\left(\begin{array}{cc}
		\frac{p_u^2}{m^2}&-1+\frac{p_up_v}{m^2}\\  
		-1\frac{p_vp_u}{m^2}&\frac{p_v^2}{m^2}
                                 \end{array}\right)_{\alpha\beta}f(p)
				 =\frac{i}{2}\left(\begin{array}{cc}
		-\frac{p_u}{p_v}&1\\  
		1&-\frac{p_v}{p_u}
                                 \end{array}\right)_{\alpha\beta}f(p).
 \ee
The second part of the equation follows by use of the mass shell condition
at $p_ap_a=0$. But this is precisely (\ref{hypPXcom})! Hence with $c=m$,
(\ref{hypvar}),
we have precisely the same solution. Using the polarization vectors of
that case we conclude that there exist two conjugate pairs in the
$\{u,v\}$-sector. \\
In the $\{2,3\}$-sector which is already diagonal we may also choose the
same polarization vectors as before and have thus one conjugate pair there.
The symmetry $SO(1,1)\times SO(2)$ is manifest. However, now the mass
being non-zero we may apply the boosts $M_{02}, M_{03}$ and the rotations
$M_{12}, M_{13}$ and realize the complete $SO(1,3)$ of the four-dimensional
Minkowski momentum space. After any one of these transformations we have
to identify the physical states as the ones obtained from the previously 
chosen states together with their transformed polarization vectors. But
this is a covariant procedure. The massless limit can, of course, not be
performed and requires the transition to a $(1,1)\times (0,2)$ spacetime
as shown above in the discussion of the case $X(<_0)$ to $Q(<_0)$.\\

\subsection{From $X=K$ to $Q(K)$}
In \cite{MuchPottelSibold} we constructed Hermitian operators $K_\mu$
as charges on Fock space forming together with translations,
Lorentz transformations and dilatations the conformal algebra.
In covariant normalization of the annihilation and creation operators
they read
 
\begin{align}\label{specconf}
   K_0=&\int{}\frac{d^3p}{2\omega_p}\omega_p~a^\dagger(\mathbf{p})
	                   \partial^l\partial_l a(\mathbf{p}) \\
   K_j=&\int{}\frac{d^3p}{2\omega_p}a^\dagger(\mathbf{p})
   	           \left(p_j\partial^l\partial_l
                    -2p^l\partial_l\partial_j-2\partial_j\right)a(\mathbf{p}) 
\end{align}
In the present subsection we inquire which operators $Q_\mu$ one can find such that (\ref{PQfock})
is satisfied. As states we use one-particle states with vanishing mass.
The operators $K$ give rise to the following variations of the
creation operator

\begin{align}\label{xconftransa1}
[K_0,a^\dagger(\mathbf{p})]=& \omega_p\partial^l\partial_l
	                             a^\dagger(\mathbf{p})\\
\label{xconftransa2}
[K_j,a^\dagger(\mathbf{p})]=&\left(p_j\partial^l\partial_l
       -2p^l\partial_l\partial_j-2\partial_j\right)a^\dagger(\mathbf{p})
\end{align}
It will turn out
that two cases have to be distinguished: in the first one the complete group 
$SO(2,4)$ is realized (as fitting to a spacetime $(1,3)$); in the second
the rotations $M_{12}, M_{13}$ and the boosts $M_{02}, M_{03}$ are not
realized; we have at disposal only the group $SO(1,1)\times SO(2)$
(as fitting to an conformal group over a $(1,1)+(0,2)$ spacetime).
We use the group names as labels for the two cases.

\subsubsection{The $SO(2,4)$ case}
We start from (\ref{xconftransa2})

\be\label{3conf}
K_j|{\bf p}>=(p_j\partial^l\partial_l-2p^l\partial_l\partial_j-2\partial_j)|{\bf p}>
\qquad j=1,2,3
\ee
We form

\be\label{3contract}
K_rP^rP_j|{\bf p}>=p_jp^r(p_r\partial^l\partial_l-2p^l\partial_l\partial_r
  -2\partial_r)|{\bf p}> \qquad r=1,2,3
 \ee
 Rewriting (\ref{3conf}) by use of (\ref{3contract}) we get

 \be\label{preQ}
 -2(p^l\partial_l+1)\partial_j|{\bf p}>=(K_j+\frac{1}{\omega^2_p}K_rP^rP_j
            +\frac{2}{\omega^2_p}p_jp^l\partial_lp^r\partial_r)|{\bf p}>
 \ee
With the identifications 

\begin{align}\label{Q3conf}
	Q_j|{\bf p}>=&i\partial_j|{\bf p}>\qquad j=1,2,3\\
	D=&i(1+p^l\partial_l)|{\bf p}>
\end{align}
We arrive at

\begin{align}\label{QjD3conf}
Q_jD|{\bf p}>=&\frac{1}{2}\left(K_j+K^r\frac{P_rP_j}{P^2_0}
                +2(D-i)^2\frac{P_j}{P^2_0}\right)|{\bf p}>\\
Q_j|{\bf p}>=&\frac{1}{2}\left(K_j+K^r\frac{P_rP_j}{P^2_0}
                +2(D-i)^2\frac{P_j}{P^2_0}\right)D^{-1}|{\bf p}>
\end{align}
An equivalent form is

\be\label{3covQconf}
Q_j|{\bf p}>=\frac{1}{2}\left(K_j-K^r\frac{P^rP_j}{P^lP_l}
                               -2(D-i)^2\frac{P_j}{P^lP_l}\right)D^{-1}|{\bf p}>
\ee
which refers to spatial components of fourvectors only and is manifestly covariant
with respect to spatial rotations.\\
The identification (\ref{Q3conf}) implies that we have conjugate pairs for the
three spatial components. It implies however also that

\be\label{P0Qjconf}
\lbrack P_0,Q_j\rbrack |{\bf p}> =-i\frac{p_j}{\omega_p}|{\bf p}>.
\ee
Lorentz covariance is definitely not manifest and the conjugation commutator
is not diagonal. The r.h.s.\ of (\ref{P0Qjconf}) would project to zero on
states carrying the projector $\eta_{jk}-P_jP_k/(P^lP_l)$. This will
require further study to follow shortly.\\

\noindent
In the next step,
when searching for a $Q_0$, we may procede in a completely analogous manner.
We start from

\be\label{0conf}
K_0 |{\bf p}> = (\omega_p \partial^l\partial_l) |{\bf p}>,
\ee
form

\be\label{preQ0}
(K_0+\frac{p^r}{\omega_p}K_r)|{\bf p}>=\frac{2}{\omega_p}p^r\partial_r(-p^l\partial_l)|{\bf p}>
\ee
and end up with

\be\label{Q0confpre}
(K_0+\frac{p^r}{\omega_p}K_r) |{\bf p}>=-2Q_0D |{\bf p}>
\ee
once we identify $D$ as usual and

\be\label{Q0conf}
Q_0 = \frac{i}{\omega_p}p^r\partial_r.
\ee
This $Q_0$ is however not Hermitian and its Hermitian part commutes with
$P$.\\

We might, of course, accept a non-Hermitian $Q_0$ and pursue the respective analysis
(we shall take up this discussion below), but for the time being
we prefer to choose $Q_0=0$ and to go along with this choice. The choice
is suggested
by two observations to be presented below in section 3.1.2. and corresponds, in the
analogy to the quantization of a massless vector field, to use Coulomb gauge: in
that context one works with a vanishing zeroth component of the vector field,
$A_0=0$, thus gives up manifest Lorentz covariance and shows afterwards that
covariance is nevertheless maintained for physical quantities.\\
With these considerations in mind we first collect the commutation relations
of $P_\mu$ with $Q_\nu$

\be\label{PQCoul}
\lbrack P_\mu,Q_\nu\rbrack|{\bf p}> \equiv iC_{\mu\nu}|{\bf p}>=
                i\left(\begin{array}{cc}
                            0&-p_k/\omega_p\\
                            0&             \\
                            0&\eta_{jk}+\frac{p_jp_k}{\omega^2_p}\\
                            0&             \\
                                   \end{array}  \right)|{\bf p}>
\ee
and then define polarization vectors \cite{BjorkenDrell}: in the given
Lorentz frame we choose
two unit vectors $\epsilon^{(\lambda)}({\bf p}), (\lambda=2,3)$ with time component zero, orthogonal to
each other and to the unit vector ${\bf p}/\omega_p$ with the orientation
${\bf p}/\omega_p={\bf \epsilon^{(2)}} \times {\bf \epsilon^{(3)}}$.
In addition we introduce a timelike unit vector $\eta=(1,0,0,0)^{T}$
 (T for transposed) with the help of which
a third independent spacelike unit polarization vector $\hat{p}$ with vanishing time component can be defined:

\begin{align}\label{Coulpol}
\epsilon^{(\lambda)}_\mu\eta^{\mu\nu}\epsilon^{(\lambda)}_\nu=-1 \quad \lambda=2,3\qquad &
\epsilon^{(2)}_\mu\eta^{\mu\nu}\epsilon^{(3)}_\nu=0\\
\frac{p_\mu}{\omega_p}\eta^{\mu\nu}\epsilon^{(\lambda)}_\nu=0 \quad \lambda=2,3 \qquad &
\hat{p}_\mu=\frac{p_\mu-(p\eta)\eta_\mu}{\sqrt{(p\eta)^2-p^2}}\\
\hat{p}_\mu\eta^{\mu\nu}\hat{p_\nu}=-1\qquad \hat{p}_\mu\eta^{\mu\nu}\epsilon^{(\lambda)}_\nu=0 \quad \lambda=2,3 \qquad &
\eta_\mu\eta^{\mu\nu}\eta_\nu=1 \quad \\
\eta_\mu\eta^{\mu\nu}\hat{p}_\nu=0 \qquad &
\eta_\mu\eta^{\mu\nu}\epsilon^{(\lambda)}\nu=0 \quad \lambda=2,3 
\end{align}
These polarization vectors satify the completeness relation

\be\label{Coulcompl}
-\eta_{\mu\nu}=\sum_{\lambda=2}^{\lambda=3}\epsilon^{(\lambda)}_\mu \epsilon^{(\lambda)}_\nu
                 + \hat{p}_\mu \hat{p}_\nu-\eta_\mu\eta_\nu.
\ee
It expresses the fact that the four vectors $\epsilon^{(\lambda)}$ with
$\lambda=2,3$
and $\epsilon^{(0)}_\mu \equiv \eta_\mu, \epsilon^{(1)}_\mu \equiv \hat{p}_\mu$ span a
four dimensional space.\\
In analogy to (\ref{PXcomspin}) we calculate now the action of the commutator (\ref{PQCoul}) on 
the states $\epsilon^{(\lambda)}_\rho|{\bf p}>$ for $\lambda = 0,...,3$.

\be\label{PQpolCstate}
iC_{\mu\nu}\eta^{\nu\rho}\epsilon^{(\lambda)}_\rho|{\bf p}>=i\left\{
\begin{array}{cl}
                0 &\qquad {\rm for} \quad\lambda = 0\\  
                 \epsilon^{(0)}_\mu &\qquad {\rm for} \quad\lambda = 1\\  
                 \epsilon^{(2)}_\mu &\qquad {\rm for} \quad\lambda = 2\\  
                 \epsilon^{(3)}_\mu &\qquad {\rm for} \quad\lambda = 3  
\end{array}\right\}|{\bf p}>
\ee
The ``scalar'' state $\lambda=0$ is mapped to zero; the ``longitudinal'' state
$\lambda=1$
is mapped onto the scalar state; the "transverse" states $\lambda=2,3$ are
diagonally mapped
onto themselves. Using as metric $\eta_{\lambda\lambda'}$ in the transverse
sector
those states have positive definite norm. On the quotient space $\{\lambda=0,1,2,3\}/\{\lambda=
0,1\}$ we have two conjugate pairs for the spatial directions two and three.\\
The completeness relation (\ref{Coulcompl}) contains information on Lorentz covariance of the
setting presented here. Since spacelike vectors remain spacelike and timelike vectors remain
timelike it is obvious that the whole state space changes under a Lorentz transformation, but
the divisor also changes and just removes the offending pieces which could introduce indefinite
metric in the transverse states. Effectively the quotient space is Lorentz covariant.\\

\noindent
This result sheds also light on the ``Lorentz gauge'':

if we were to use a non-Hermitian $Q_0$
we could introduce manifestly Lorentz covariant polarization vectors, but due to the
non-Hermitian nature of $Q_0$ we also had to form a quotient space which would then be just
equivalent to the Coulomb gauge case.\\
As to locality a similar comment applies as in the case of $Q(\nabla)$.
Although $K$ is local in $x$-space, $Q$ has to be generated from it
by ``dividing'' through $D$. And this is certainly a non-local operation
(cp.\ equation (103) in \cite{EdenSibold}).

\noindent
The solution for general $n$-particle states has to be constructed via symmetrized tensor
products. We do not go into details of this problem.\\

\subsubsection{The $SO(1,1)+S(0,2)$ case}
\noindent
The conformal algebra can also be represented in a form which is closely
related to the symmetry which governed the $<_0$ case:
$SO(1,1)\times SO(2)$.
Here two boosts and two rotations are trivially represented. One may interpret
this type of model as being fully realized on four dimensional spacetime
with standard representation of the Lorentz group for all quantities but
the (``would-be'') observables $X$, resp.\ $Q$. Alternatively one can
interpret the underlying spacetime to be $(1,1)\times(0,2)$ and the full
algebra of it to be implemented. 
In any of the two interpretations we have to restrict the generators and
relabel the states accordingly if we wish to realize this algebra correctly
on suitable one-particle Fock states. For the states we shall write

\be\label{sttensor}
|{\bf p}>=|p_1>|p_a>\equiv|p_1;p_a> \qquad a=2,3.
\ee
For the algebra we introduce

\[\begin{array}{rclrcl}\label{splitalgebra}
P_0|>&=&\omega_p|>& P_2|>&=& p_2|>\\
P_1|>&=&p_1|> & P_3|>&=&p_3|>\\
M_{10}|>&=&i\omega_p\partial_1|>& M_{23}|>&=&
	       -i(p_2\partial_3-p_3\partial_2)|>\\
M_{01}|>&=&-M_{10}|>& M_{32}|>&=&-M_{23}|>\\
D^{(1,1)}|>&=&ip^1\partial_1|>&
D^{(0,2)}|>&=&i(1+p^2\partial_2+p^3\partial_3)|>\\
K_0|>&=&\omega_p\partial^1\partial_1|>&
K_2|>&=&(p_2\partial^b\partial_b-2(p^b\partial_b
	       +1)\partial_2)|>\\
K_1|>&=&-p_1\partial^1\partial_1|>&
K_3|>&=&(p_3\partial^b\partial_b-2(p^b\partial_b+1)\partial_3)|>\\
\multicolumn{3}{l}{(\omega_p\equiv\sqrt{-p^1p_1})}
&\multicolumn{2}{l}{|>\equiv |p_1;p_a>}
 \end{array}\]
Hence on the factors $|p_1>$, resp.\ $|p_2,p_3>$, the conformal algebras for
spacetimes with one time $+$ one space dimension, $(1,1)$, resp.\ zero time $+$
two space dimensions, $(0,2)$, are realized.\\
The boosts $M_{20},M_{30}$ and the rotations $M_{12}, M_{13}$ of the ambient
spacetime
$(1,3)$ with conformal group $SO(2,4)$ are not realized; they correspond to
those Lorentz transformations whose massless limit did not exist and had to
be discarded there.\\
Turning our attention now to the construction of $Q$ we first observe that
on the purely spatial part $(0,2)$ we have identical formulas as compared with
the previous case $(1,3)$, the range of the indices being restricted to
$a=2,3$.
Hence we have identical results: The operators $Q_a, a=2,3$ are given by

\be\label{so2QK}
Q_a|p_1;p_a> = \left[K_a-(D-i)^2\frac{P_a}{P^bP_b}\right]D^{-1}|p_1;p_a>
\ee
where the range of $b$ (summation) is also $2,3$ and $D\equiv D^{(0,2)}$.
They have the canonical form

\be\label{so2canQ}
Q_a|p_1;p_a>=i\partial_a|p_1;p_a> \qquad a = 2,3.
\ee
Again we have to have a look to the fate of the commutator $\lbrack P_0,Q_j\rbrack$.
That it is indeed vanishing in the present situation can be checked when using the
 full expression (\ref{so2QK}), e.g.\ on the state $P^bP_b D|p_1;p_a>$.\\
In the $(1,1)$ part we note that the $D^{(1,1)}$ and the $M_{10}$ 
as well as the $K_0$ and $K_1$ transformations at most differ by a sign from
each other.
This implies on the one hand that the $M$-contribution in the commutator
$[P,K]$ is simply related to the $D$-contribution and on the other hand
that we can avoid using the projector $P/PP$ contracted with $K$.
Indeed

\be\label{conf11algebr}
[P_\alpha,K_\beta]|p_1;p_a>=2\left(\begin{array}{cc}
	                         -p^1\partial_1&\omega_p\partial_1\\
		                  -\omega_p\partial_1&p^1\partial_1
\end{array}\right)_{\alpha\beta}|p_1;p_a> \qquad \alpha, \beta=0,1
\ee 
Hence on $\frac{1}{2}D^{-1}|p_1;p_a>$ (note: $D$ commutes with $[P,K]$)

\be\label{D-1conf11algebr}
[P_\alpha,K_\beta]\frac{1}{2}D^{-1}|p_1;p_a>=i\left(\begin{array}{cc}
	                         1&\varepsilon\\
		                  -\varepsilon&-1
\end{array}\right)|p_1;p_a>_{\alpha\beta} 
	  \qquad \alpha, \beta=0,1; \varepsilon=\frac{p_1}{\omega_p}=\pm1
\ee
In order to diagonalize the system we introduce

\be\label{PKplusminus}
P^{(\pm)}=\frac{1}{2}(P_1\pm P_0) \qquad K^{(\pm)}=\frac{1}{2}(K_0\pm K_1)
\ee
and then find

\begin{align}\label{PQplusminus}
 [P^{(+)},K^{(-)}]\frac{1}{2}D^{-1}|p_1;p_a>=&+i|p_1;p_a> \qquad  \varepsilon=+1\\
	[P^{(-)},K^{(+)}]\frac{1}{2}D^{-1}|p_1;p_a>=&-i|p_1;p_a> \qquad  \varepsilon=-1,
\end{align}
whereas the other commutator entries vanish. In matrix form this reads

\be\label{PQeta}
[P^{(\pm)},K^{(\mp)}]\frac{1}{2}D^{-1}|p_1;p_a>
        = i\left(\begin{array}{cc}
	        	1&0\\
		        0&-1
	\end{array} \right)|p_1;p_a>=i(\eta)_{\alpha\beta}|p_1;p_a> \quad \alpha, \beta=+,- 
\ee
It is thus legitimate to interpret the operator on the l.h.s.\ as the commutator
of a conjugate pair $P,Q$. The r.h.s.\ tells one that the norms of the states generated
by this pair are opposite in sign, hence the best one can do is to prescribe
a kind of Gupta-Bleuler condition by requiring that the physical states must
always contain an equal number of $[P^{(+)},K^{(-)}]$ factors. The relation
is covariant under application of Lorentz boosts in the $(0,1)$-plane, i.e.
the boost belonging to the little group of $SO(1,3)$, since
$P_\alpha, K_\beta$ are vector operators w.r.t.\ $SO(1,1)$ and $D^{(1,1)}$
commutes with $M_{\gamma,\delta}$ ($\alpha, \beta, \delta, \gamma =0,1$).
The $SO(2)$-factor is not
touched by these transformations and is itself covariant under the
$SO(2)$-transformations.\\

\noindent
Again, for general $n$ one has to construct tensor products.\\

\newpage

\section{Group theoretic approach}  

\noindent
The construction of conjugate pairs of operators in relativistic QFT has in 
particular been pursued by using group theoretic methods. In \cite{LaguLaue}
it has been based on the algebra of the conformal group $SO(2,4)$
interpreted as acting on four dimensional Minkowski spacetime. In the
first subsection we review this work to some extent and thereafter
put it into perspective of our present paper.\\

\subsection{Representation of the conformal group including $Q$}
In \cite{LaguLaue} a representation of the conformal algebra has been
established by going over to the enveloping algebra, where the
standard generators $\{P_\mu,M_{\mu\nu},D,K_\mu\}$, translations,
Lorentz transformations, dilatations, special conformal transformations 
respectively had been replaced by $\{P_\mu, S_{\mu\nu}, Y, Q_\mu\}$ such
that $P$ and $Q$ form a conjugate pair and operate on a Hilbert space $H_Q$;
$S$ satifies commutation relations with itself like $M$ does, represented
on a Hilbert space $H_S$, whereas the single $Y$ generates an irreducible,
hence
one dimensional representation on a Hilbert space $H_Y$. Assuming that
these three Hilbert spaces are different the representation is based on
the tensor product $H_Q \otimes H_S \otimes H_Y$.\\
In our notations and conventions
one starts with some Hilbert space of functions of one variable and defines
on it differential operators $P,Q$ which satisfy

\be\label{LLPQcom}
[P_\mu,Q_\nu]=i\eta_{\mu\nu}\qquad [P_\mu,P_\nu]=0 \qquad [Q_\mu,Q_\nu]=0
\ee
Next one introduces operators 

\begin{align}\label{LLconf}
	M_{\mu\nu}= & Q_\mu P_\nu-Q_\nu P_\mu + S_{\mu\nu}\\	
	D= & \frac{1}{2}(PQ+QP) + Y= QP+2i+Y\equiv QP+Y'\\
K_\mu =& 2(-Q_\mu QP+\frac{1}{2}Q^2P_\mu+Q_\mu D+Q^\lambda M_{\mu\lambda})
\end{align}
with $S_{\mu\nu}=-S_{\nu\mu}$. One can convince oneself that the new set of
operators \{P,Q,S,Y\} closes once one assumes that $Y$ commutes with $P,Q,S$.
The aim is now to express $Q,Y,S$ in terms of the original operators
\{P,M,D,K\}.
In \cite{LaguLaue} it has been shown that $Y,S$ can be expressed in terms of the
 Casimir operators of the conformal group. This information has then been used
 first for giving an interpretation of these Casimir operators as
 {\sl conformal spin} for $S$, as {\sl fundamental length} for $Y'$; second 
 for discussing  the irreducibility  of this new representation of the conformal 
 group. Of particular importance is  the inversion for $Q$. It is performed via
 combination

\be\label{KQproj} 
\frac{1}{2}K^\lambda(\eta_{\lambda\mu}-\frac{P_\lambda P_\mu}{P^2})
  = Q^\lambda Y'(\eta_{\mu\lambda}-\frac{P_\mu P_\lambda}{P^2})
      + Q^\lambda M_{\rho\lambda}(\eta^\rho_\mu - \frac{P^\rho P_\mu}{P^2})
      \ee
Here the expression for $D$ and the commutator (\ref{LLPQcom}) have been used and $Y$
      has been replaced by $Y'$. Clearly, this formula makes sense only if $P^2$ does
      not vanish. In \cite{LaguLaue} it has been  argued by counting number
      of unknowns and number of 
      equations that one can solve for $Q$. We note however and discuss in more detail below that $K$ and $Q$
      are contracted with the transverse projection operator $\eta_{\mu\nu} -P^\mu P^\nu/P^2$, hence their
      relation might be determined only up to a longitudinal term, proportional to $P/P^2$.\\
In the case $S=0$ one inserts the expression for $M$ in terms of $P,Q$, uses also 
$D$ as function of $P,Q$ and arrives at

\be\label{QLLofK}
Q_\mu=\left[\frac{1}{2}K^\lambda(\eta_{\mu\lambda}-\frac{P_\mu P_\lambda}{P^2}) 
              +D(D-2Y-4i)\frac{P_\mu}{P^2}\right]D^{-1}
\ee
I.e.\ In this case a suitable longitudinal term showed up and the solution
is unique.\\
Returning to the general case, $S\not=0$ one notes that a representation of
(\ref{LLPQcom}) on a Hilbert space $H_Q$ the latter
must contain at least square integrable functions $f(p)$, the scalar product
being given by $(f,g) = \int_{V_{+}}d^4pf^{*}(p)g(p)$ with $V_{+}$ denoting
the forward cone of $p^2>0$. On this domain $P_\mu$ is self-adjoint and
the $Q_\mu$'s are given by $i\partial/\partial p_\mu$ which is Hermitian
but not self-adjoint. Their domain of Hermiticity is the dense set
of differentiable functions of $p_\mu$ which vanish on the boundary of $V_{+}$.
The operators $K$ are self-adjoint on $H_Q$: an irreducible representation
for the conformal group has been found and the Casimir invariants are 
multiples of the identity.\\
Other, equivalent representations are given by functions which have support
either for spacelike $p_\mu$, i.e.\ $p^2<0$, or lightlike $p_\mu$, i.e.\
$p^2=0$, or the negative cone $V^{-}=\{p\in{\mathbb{R}}^4|p^2>0,p_0<0\}$.
But due to the fact that a self-adjoint $Q_\mu$ has its spectrum on the 
entire line, the decomposition into several irreducible representations
does only yield Hermitian $Q_\mu$.\\

\subsubsection{Non-commutative coordinates}
Having with (\ref{QLLofK}) at hand an operator $Q$ which forms together
with $P$ a conjugate pair we can realize a {\sl non-commutative coordinate
operator} via

\be\label{nccoordk}
Q^{\rm nc}_\mu = Q_\mu + \Theta_{\mu\nu}P^\nu
\ee
with $\Theta$ real and anti-symmetric. $Q^{\rm nc}$ clearly satisfies

\be\label{nccomco}
[Q^{\rm nc}_\mu,Q^{\rm nc}_\nu]=2i\Theta_{\mu\nu}
\ee
(cp.\ (\ref{nccoord}).\\
The definition of $Q^{\rm nc}$ and the commutation relation (\ref{nccomco}) 
hold on the function space described before for $Q$ and $P$ and likewise 
they have the same domain of Hermiticity. We leave open the question in which
sense these properties indeed qualify $Q^{\rm nc}$ as a ``true'' 
non-commutative coordinate operator.\\
We note however that restricting the functions $f$ on which $Q^{\rm nc}$ acts
to obey equations of motion, i.e.\ to go on-shell, one will encounter the
intricacies which have been presented for $Q=Q(K)$ in subsect.\ 2.3.
These will be discussed now.\\

\subsubsection{Consistency of off-shell/on-shell treatment for $S=0$}
The above considerations hold on a Hilbert space of functions $f(p)$ which
do not necessarily satisfy any differential equation. In the parlance of QFT
one could understand them as off-shell one-particle Green functions. The
considerations of section 2 refer to one-particle states, i.e.\
wave functions solving the respective Klein-Gordon equation. It is
then natural to inquire how the results of the preceding subsection
are related to them.\\
As first topic we show how our variations of one-particle states with respect
to $K$ (\ref{xconftransa1},\ref{xconftransa2}) come out from (\ref{LLconf}).
$K$ has been defined as

\be \label{LLKconf}
K_\mu = 2(-Q_\mu QP+\frac{1}{2}Q^2P_\mu+Q_\mu D+Q^\lambda M_{\mu\lambda}),
\ee
We interpret now the operators as differential operators $\delta^A (A=P,M,D)$ 
acting on some eigenfunction of $P$, hence obtain in the first step

\be\label{deltaKofQLL}
\delta^K_\mu=2(-p^\nu\delta^Q_\nu\delta^Q_\mu
              +\frac{1}{2}p_\mu\delta^Q_\lambda\delta^\lambda_Q
              +\delta^D\delta^Q_\mu+\delta^M_{\mu\nu}\delta^\nu_Q)
\ee
Eventually we wish to realize $Q_j$ by $i\partial/\partial p^j$ and
therefore use as $\delta$'s for $A=D,M$ our standard variations and 
find in the second step

\begin{align}\label{deltaKofQLL2}
\delta^K_0|{\bf p}>=&(2i\delta^Q_0
          +\omega_p\frac{\partial^2}{\partial p^l\partial p_l})|{\bf p}>\\
\delta^K_j|{\bf p}>=&(-2i\omega_p\delta^Q_0\frac{\partial}{\partial p^j}
		   +2i\omega_p\frac{\partial}{\partial p^j}\delta^Q_0
		   +p_j\frac{\partial^2}{\partial p^l\partial p_l}
		   -2p^l\frac{\partial^2}{\partial p^l\partial p^j}
		   -2\frac{\partial}{\partial p^j}|{\bf p}>.
\end{align}
This result tells us that for $\delta^Q_0=\partial/\partial p^0$
the construction within \cite{LaguLaue} provides a relation between all variations $\delta^K$
and all variations $\delta^Q$ which as we know from the $S=0$-case
one is able to invert.
For $\delta^Q_0=0$ in the relation for $\delta^K_0$,
i.e.\ no independent variation with respect to direction $0$, i.e.\ 
$\partial/\partial p^0\equiv 0$, we obtain precisely our on-shell variations
$\delta^K$. Hence we conclude that the two approaches match.\\

\noindent
As the second topic we discuss -- for the \cite{LaguLaue}-case $S=0$ and a 
representation with $P^2=0$ -- what we shall call the ``gauge'' problem.\\
We use the solution (\ref{QLLofK}) and apply it to a
one-particle state $DP^2|{\bf p}>$

\begin{align}\label{Q0ops}
	(Q_0DP^2)|{\bf p}>= &(\frac{1}{2}\times 0
           -\frac{1}{2}\omega_p p^\lambda\delta^K_\lambda
	   +\omega_p (i(1+p^l\partial_l)-4i-2y)i(1+p^l\partial_l))|{\bf p}>\\
	   =& 0 \qquad {\rm for}\quad y=-i
   \end{align}
The ``direct'' term $K_0$ is annihilated by $p^2=0$ (on-shell-ness);
however the projector contribution $K^\lambda P_\lambda P_0$ is
{\sl non-trivially} cancelled by 
the contribution coming from the $D$-terms. For $\mu=j$ however no cancellation
takes place and we arrive at a contradiction: the l.h.s.\ vanishes, the
r.h.s\ does not. 
Hence, like in the quantization of (massless) gauge fields we have to give up at least one of the fundamental
properties which we would have liked to be realized. In section 2.3.1 we
gave up manifest Lorentz covariance, used $Q_0=0$ (Coulomb gauge) and were
able to realize two conjugate
pairs on states with definite metric. If we had sticked to manifest
covariance we would have had to give up Hermiticity for $Q_0$.\\
We shall see in the next subsection that a similar phenomenon happens in the
massive case.\\

\subsection{Representation of Poincar\'e and dilatations}
For $n=1$ the relation (\ref{PXnablacom}) can be rewritten as
 
\be\label{PXnablacompos}
[P_\mu,X^{\nabla}_\nu]=i(\eta_{\mu\nu}-\frac{P_\mu P_\nu}{P^2})
\ee
It is then suggestive to introduce an operator $X^{({\rm com})}$ (``com'' for 
``composite'')

\be\label{Xcomp} 
X^{({\rm com})}_\mu=M_{\mu\lambda}\frac{P^\lambda}{P^2}
\ee
which is a Lorentz vector

\be\label{XcompLor}
[M_{\mu\nu},X^{({\rm com})}_\rho]= 
-i(\eta_{\mu\rho}X^{({\rm com})}_\nu-\eta_{\nu\rho}X^{({\rm com})}_\mu),
\ee
fulfils

\be\label{XcompXcomp}
[X^{({\rm com})}_\mu,X^{({\rm com})}_\nu]=iM_{\mu\nu}\frac{1}{P^2}.
\ee
i.e.\ the analogue of (\ref{XXcom}), and reproduces (\ref{PXnablacompos}).\\
We now choose eigenfunctions of $P$ as representation space, interpret
the operators involved accordingly as differential operators and 
apply $X^{({\rm com})}$ to an eigenfunction $\phi(p)$

\be\label{Xcomp1off}
X_\mu^{({\rm com})}\phi(p)=
    i\left(\frac{\partial}{\partial p^\mu}
     -\frac{p_\mu}{p^2}p^\lambda\partial_\lambda \right)\phi(p)	     
\ee
The first derivative term points to an operator $Q$ which indeed is
realized once we add a term $(D-\hat{Y})(P_\mu/P^2)$ with
$D=i(1+p^\lambda\partial p^\lambda), \hat{Y}=i$ on $\phi(p)$.        
 Hence

\be\label{Qcompnablaoff}
Q_\mu=X^{({\rm com})}_\mu+(D-\hat{Y})\frac{P_\mu}{P^2}
\ee
yields 

\be\label{nablacomconjoff}
Q_\mu\phi(p)=i\frac{\partial}{\partial p^\mu}\phi(p)
\ee
We note first of all that adding $(D-\hat{Y})P_\mu/P^2$ to $X^{({\rm com})}$
generates an
{\sl abelian} operator $Q_\mu$ (four components !), second that
$Q_\mu$ obviously satisfies the conjugation relation
$[P_\mu,Q_\nu]=i\eta_{\mu\nu}$ on the eigenfunctions $\phi(p)$.\\
We therefore succeeded to find an operator (in the enveloping algebra of
Poincar\'e + dilatations) which realizes $Q_\mu=i\partial/\partial p^\mu$.
It is also noteworthy that the differential operator on the r.h.s.\ of
(\ref{Xcomp1off}) is just an off-shell continuation of $\nabla_\mu$.\\

\subsubsection{Non-commutative coordinates}
In perfect analogy to the conformal case we are also in the present context able
to define a differential operator which qualifies -- at least formally -- as a
{\sl non-commutative coordinate operator}\ :

\be\label{massnccoord}
Q^{nc}_\mu = Q_\mu + \Theta_{\mu\nu}P^\nu 
= X^{({\rm com})}_\mu + (D-\hat{Y})\frac{P_\mu}{P^2}+ \Theta_{\mu\nu}P^\nu
\ee
(Again, $\Theta$ is real and antisymmetric.)
It operates on functions $\phi(p)$, with $P,M,D$ accordingly interpreted as
differential operators. It is to be noted that the mass can be either
non-vanishing or (for off-shell $\phi$) vanishing .\\

\subsubsection{Consistency of off-shell/on-shell treatment for $S=0$}
Let us now choose Fock space as representation space. Then formulae exactly
analogous to the above ones hold on one-particle states with range of indices
$\lambda$ restricted to \{1,2,3\}:

\be\label{Xcomp1}
X_\mu^{({\rm com})}|{\bf p}>=
     i\left(\frac{\partial}{\partial p^\mu}
	     -\frac{p_\mu}{m^2}p^l\partial_l \right)|{\bf p}>	     
\ee

\be\label{Qcompnabla}
Q_\mu=X^{({\rm com})}_\mu+(D-\hat{Y})\frac{P_\mu}{P^2}
\ee

\be\label{nablacomconj}
Q_\mu|{\bf p}>=i\frac{\partial}{\partial p_\mu}|{\bf p}>
\ee
We obtain $Q_0=0$ once we put the derivative
$\partial/\partial p^0 \equiv 0$. Thus these considerations confirm on the one
hand that one can invert off-shell, on the other hand that 
our on-shell arguments on the vanishing of $Q_0$ in subsubsction 2.1.1 are
correct. \\

\noindent
Obviously the above formulae are very close to those of \cite{LaguLaue}
for an operator $Q_\mu$ derived from the conformal generators $K_\mu$.
The precise derivation proceeds as follows. We use the definitions of
(\ref{LLconf}) for $M$ and $D$ obtain  

\begin{align}\label{QLLDM}
	M_{\mu\lambda}P^\lambda=& Q_\mu P^2-QPP_\mu+S_{\mu\lambda}P^\lambda\\
	D=&QP+2i+Y\\
	Q_\mu=&\frac{M_{\mu\lambda}P^\lambda}{P^2}
	+\left((D-(Y+2i))\eta_{\mu\lambda} -S_{\mu\lambda}\right)\frac{P^\lambda}{P^2}\\
		=&X^{({\rm com})}_\mu
+\left((D-(Y+2i))\eta_{\mu\lambda}-S_{\mu\lambda}\right)\frac{P^\lambda}{P^2}
\end{align}
For $S=0, \hat{Y}=Y+2i$ this is precisely our expression (\ref{Qcompnabla}).
The only difference is, that in our ad hoc approach the abelian character
of $Q$ comes out as a result, whereas here, going along the lines of
\cite{LaguLaue}, it has been assumed from the start. But clearly, the
main content is the same.\\

\noindent
In the vein of the present section ``group theoretic approach'' these 
considerations can be interpreted as the fact, that the operators $\{P,Q,Y\}$
generate the same representation of the group
${\hbox{\rm Poincar\'{e}}} \times {\rm dilatation}$
as the set of generators $\{P,M,D\}$ via the identification (\ref{LLconf})
with $S=0, Y=i$.\\

\noindent
Finding one and the same $Q$ on Fock space starting from different expressions
in different algebras is just analogous to the well-known fact in
QFT \`a la Lehmann-Symanzik-Zimmermann, that different interpolating fields
may represent one and the same particle on-shell.\\

\newpage

\section{Discussion, conclusions, open questions}
\subsection{Universality}

We first summarize our findings schematically in a table:\\

\vspace{-20pt}


\small

\[
\begin{array}[t]{ccccccc}
\multicolumn{7}{l}{\rm{\hbox{Table: cases of preconjugate and conjugate
variables}}}\\
\hline
  &{\rm prec.}
  & 
  &{\rm conj.}
  &\begin{array}{c}
	           {\rm symm.\ of}\\
	           {\rm spacetime}
   \end{array}		
  &\begin{array}{c}
                   {\rm state\ space\ type}\\
                   {\rm state\ space\ symm.}
		        \end{array}
  & \begin{array}{l}
          {\rm number\,\, of}\\
          {\rm conj.\ pairs}
    \end{array}\\
\hline    
m^2\not=0\quad
  &X(\nabla)
  &\rightarrow
  &Q(\nabla)&SO(1,3)
  &\begin{array}{c}
        {\rm standard}\\
      	SO(1,3)
    \end{array}
  &3\,\, {\rm spatial}\\
\hline    
m^2\rightarrow 0\, 
  &  
  &\rightarrow  
  &Q(\nabla)_{|m^2=0}&SO(1,3) 
  &\begin{array}{c}
        {\rm standard}\\
        SO(1,3)
   \end{array}
  &2\,\, {\rm spatial}\\
\hline   
 m^2=0\,
 &X(K)
 &\rightarrow
 & \begin{array}{c}
	  Q(K)\\
	  Q(K)
   \end{array}
 & \begin{array}{c}
	       SO(2,4)\\
	       SO(1,1)\times SO(2)	
    \end{array}
 &{\begin{array}{c}
       { \begin{array}{c}
               {\rm quotient}\\
	        SO(1,3)
          \end{array}}\\
       {\begin{array}{c} 
	      {\rm quotient}\\
	      SO(1,1)\times SO(2)
	  \end{array}}
   \end{array}}
 &\begin{array}{c}
       2\, {\rm spatial}\\
       2\, {\rm spatial}
   \end{array}\\
\hline  
m^2=0\,
 &X(<_0)
 &\rightarrow
 &Q(<_0)
 &SO(1,1)\times SO(2)
 &\begin{array}{c}
      {\rm standard}\\
      SO(1,1)\times SO(2)
   \end{array}
 &2\, +\, 1
\\
\hline
m^2\not=0\quad
  &X(<)
  &\rightarrow
  &Q(<)
  &SO(1,3)
  &\begin{array}{c}
        {\rm standard}\\
      	SO(1,3)
    \end{array}
  &2 + 1
\end{array}    
\]    

\normalsize

\noindent
and then describe them in detail.\\
In the massive case we started from 
$X(\nabla)$, s.\ (\ref{Xnabla}), which has geometrical meaning,
and then derived the on-shell quantities $Q(\nabla)$, s.\ (\ref{Q0Qj}).
Here it is crucial to rely on the presence of polarization vectors.
The fact that $Q^{({\rm eff})}_0=0$ can however be seen already
when looking at the off-shell quantities \cite{LaguLaue}-type
$Q_\mu$, s.\ (\ref{Qcompnablaoff}), which originate from group theoretic
considerations. Going on-shell there confirms the vanishing of
$Q(\nabla)_0$. Universality clearly means ``equality on Fock
space'' which obviously has been achieved. Three (spatial) conjugate
pairs exist. Due to the polarization vectors they operate on states
with positive definite norm. Lorentz covariance is non-manifestly 
realized.\\
In the limit of vanishing mass this
structure of physical state space can be maintained, but $Q_1$
vanishes, hence only two spatial pairs survive.\\
The generically massless case has been based on the preconjugate
$X_\mu=K_\mu$
(\ref{specconf}), $K$ generating the special conformal transformations.
The version relevant for this universality sector is based on the
spacetime with dimension $(1,3)$. Here also $Q(K)_0=0$, confirmed
via off-shell reasoning, (\ref{Q0ops}), and -- in order to diagonalize
the conjugation commutator -- one has to mode out one spatial component.
Two spatial conjugate pairs exist.\\
Quite natural, however, seems to be a truncation of the algebra to
$SO(1,1)\times SO(2)$ and the spacetime to be $(1,1)+(0,2)$. On the
state space (\ref{sttensor})
we found two conjugate pairs for the spatial part $(0,2)$; those 
over the $(1,1)$ part have to be moded out for norm reasons.\\
A class of special interest is formed by $X(<_0)$ with its associated
operator $Q(<_0)$. In the massless limit (of $X(<)$ to $X(<_0)$)
the symmetry shrinks to
$SO(1,1)\times SO(2)$ and accordingly also the spacetime to
$(1,1)+(0,2)$. Since however in momentum space a {\sl double} cone is
realized as opposed to the {\sl single} (forward) cone in the previous
examples $(Q(\nabla), Q(K))$ the resulting outcome for $Q(<_0)$
and the state space differs from the analogous conformal case: on
the $(0,2)$ part of spacetime
one independent conjugate pair is realized on two states with
polarization vectors $\epsilon(r), r=2,3$, s.\ (\ref{PQlwso2}).
In the $(1,1)$ part of spacetime which appears however as $(u,v)$
and as $(p_u,p_v)$ on momentum space we have two conjugate pairs
operating on two states with positive definite norm. Due to the
non-diagonal form of the metric the operators $Q^u_{\rm eff},
Q^v_{\rm eff}$ have a non-vanishing commutator, (\ref{Lorgen}).\\
Once this structure has been found one can establish exactly the
same one also for non-vanishing mass, $X(<)\rightarrow Q(<)$
s.\ (\ref{mrPXcom}), and -- just due to the non-zero
mass -- one can extend it to the full Lorentz group . The number and type
of conjugate pairs coincides with the massless case and thus reaches
the maximal number obtainable: two in the $\{u,v\}$-sector,
one in the $\{2,3\}$-sector. The relevant state space is the standard
Fock space augmented by the polarization vectors.\\
An intriguing result of
our analysis may therefore be that wedge-local quantum field theories
just provide by definition the right balance between position and
momentum variables on the quantum field theoretic level to form
respective operators which come as conjugate pairs on-shell. Time
does not play a preferred role any more.\\
In order to find a direct relation between $Q(<_0)$ on the one hand,
the massless limit of $Q(\nabla)$ and $Q(K) (1,3)$ on the other
we first recall that $Q(\nabla)_0=Q(\nabla)_1=0$ in the massless limit,
s.\ (\ref{conjeqnabla2}), and that $Q(K)_0$ and $Q(K)_1$ are moded out
in the relevant state space (s. subsubsection 2.3.1).
Let us consider the {\sl quadruple}
$\{Q(<_0)_{u,v},\epsilon_\alpha^{(u,v)},A({\bf p})|0>,<0|A({\bf p})\}$
and compare it with the corresponding
quadruples
$\{Q(K)_{0,1},\epsilon_\alpha^{(0,1)},a^\dagger({\bf p})|0>,<0|a({\bf p})\}$,
$\{Q(\nabla)_{0,1},\epsilon_\alpha^{(0,1)},
a^\dagger({\bf p})|0>,<0|a({\bf p})\}.$ (The writing
should indicate that due to $A^\dagger({\bf p})=-A({\bf p})$, (\ref{real}),
as opposed to $(a^\dagger({\bf p})|0>)^\dagger = <0| a({\bf p})$ the
$Q(<_0)$ lives in a bigger space than the other two $Q$'s.)
Now, it becomes clear that the latter two are effectively the projection
to zero
of the first one (refering to the $Q$'s). The reason for the non-triviality
of $Q(<_0)_{u,v}$ is the presence of the {\sl double} cone; the reason for
the triviality of the corresponding components of $Q(K)$ and $Q(\nabla)$
(massless limit) the non-existence of $\nabla^u, resp.\ \nabla^v$ as expressed
in equations (\ref{limwedgediff}) which prohibits a $1\leftrightarrow 1$
relation.

\subsection{The gauge problem}
In the course of our investigations it has become clear that the
postulate $[P_\mu,Q_\nu]=i\eta_{\mu\nu}$ has first of all to be understood
in a weak sense: as applied to spaces of functions or states. It further
became clear that the
r.h.s.\ of the commutator equation may be interpreted like in gauge theories: 
The ``pure'' $\eta_{\mu\nu}$ form corresponds to Lorentz gauge and is
naturally realized off-shell: in the ad hoc version as Fourier transform
(no realization of $Q$ as function of other operators of the theory),
in the \cite{LaguLaue}-version $Q=Q(K)$ and in the \cite{LaguLaue}-type
construction in subsection 3.2. On-shell, i.e.\ on Fock states,
we met the Landau gauge in $X(\nabla) \rightarrow Q(\nabla)$, massive
version; the Coulomb gauge in $X(K) \rightarrow Q(K), (1,3)-{\rm spacetime}$;
light cone gauge in $X(<_0) \rightarrow Q(<_0)$.
In hindsight the explanation is simple:
the desired $\eta_{\mu\nu}$ can be expanded into a sum over polarization
vectors $-\eta_{\mu\nu}=\sum_{\lambda=0}^{\lambda=3}\epsilon^{(\lambda)}_\mu
\epsilon^{(\lambda)}_\nu$, the polarization vectors provide a basis
for the space spanned by $\eta_{\mu\nu}$, hence one is lead to define
new states $\epsilon^{(\lambda)}_\mu |{\bf p}>$.
It is then non-trivial, but true that on these states the inversion from
a preconjugate $X$ to a conjugate $Q$ is possible. The different signs
within $\eta_{\mu\nu}$ determine the norm of the eventual state.
The solution $Q_\mu=i\partial/\partial p^\mu$ on these states leads to
$Q_0=0$, since on shell no independent motion in direction zero, driven
by $\partial/\partial p^0$, is generated. $Q_0$ is however a tentative
time operator. Pauli's theorem is refined in a very bold
sense: $Q_0$ is not only not self-adjoint -- it vanishes! This must
not be understood as a surprise, after all. On-shell states are
constructed within the limit of $\pm$infinite time, hence do not move
in the flow of time. They can not serve as direct instrument to measure
time.\\
In the context of the case $Q(<_0)$ the gauge nature of the definition of
conjugate pairs points to a possible relation with the construction
of gauge theories in non-commutative field theories,
notably \cite{Wess}. This aspect remains to be explored.\\

\subsection{General fields, more general states}
Obviously fields and states carrying spin should be studied along the lines
presented in this paper. The LaguLaue construction,\cite{LaguLaue} of a
$Q(K)$ for
non-vanishing $S$ could serve as guide line and would have to be
explicitly implemented. Supersymmetry might be a helpful tool since
there the superconformal algebra spans all spacetime symmetries of the
respective theory.\\
For the construction of conjugate pairs we introduced polarization vectors
multiplying ordinary Fock states. They solved the gauge i.e.\ 
the norm problem associated with conjugate pairs. Hence these polarization
vectors should be considered as a new, essential attribute for constructing
the observables $Q$. They may be interpreted as tensoring the state space
with some factor. But this factor is in our derivation not arbitrary.
This might be in contrast with \cite{BahnsDoplicherMorsellaPiacitelli}.\\
The quadruples
$\{Q(<_0), \epsilon, A({\bf p})|0>, <0|A({\bf p})\}$, 
~$\{Q(\nabla), \epsilon, a^\dagger({\bf p})|0>~, <0|a({\bf p})\}$ and 
$\{Q(K), \epsilon, a^\dagger({\bf p})|0>, <0|a({\bf p})\}$ serve as
``detectors''
in the one-particle states of Fock space for determining the value of $Q$.\\
On a formal level these ``dressed'' states are
asymptotic w.r.t. their spacetime variables, a deeper understanding of
them would however be desirable. The inherent non-locality in $x$-space
when deriving the $Q$'s from the $X$'s and taking into account the
effect of the polarization vectors seems to be in accordance with
\cite{Yngvason}.\\
Even off-shell one could probably introduce analogous quantities and
discuss in these terms the domain questions of the operators $Q$ which
would then be related to norm properties as well.\\ 
A link should also be found to thermal states (s.\cite{VerchGranseePinamonti})
and thermal quantum fields (s.\cite{BuchholzBros}).\\

From a very general point of view it is obvious that quite a few notions
of time exist. One of them is associated with irreversible processes
giving rise to an arrow in time. Realizing something like this in  
relativistic systems requires generalization of entropy and other
thermodynamic quantities and the introduction of respective state spaces.
In the general relativistic context this might provide even more insight
and explain phenomena not understood today.\\

\noindent
Acknowledgements\\
S.P. gratefully acknowledges financial support by the Max Planck Institute
for Mathematics in the Sciences and its International Max Planck Research
School (IMPRS) ``Mathematics in the Sciences''. He would like to
thank Prof. Sergio Doplicher for helpful discussions. K.S. is grateful
for enlightening discussions with Rainer Verch and Erhard Seiler.\\

\bibliographystyle{phjcp}
\bibliography{bphzl}
\end{document}